\documentclass[aps,prl,reprint,showpacs,superscriptaddress,longbibliography,nourl,noeprint]{revtex4-2}
\usepackage{amsmath,amssymb,physics,bm}
\usepackage{graphicx, microtype,siunitx} 
\usepackage[colorlinks=true,linkcolor=blue,citecolor=blue,urlcolor=blue]{hyperref}
\usepackage[all]{hypcap} 

\newcommand{\rvec}[0]{{\bm r}} 
\newcommand{\nvec}[0]{{\bm n}}
\newcommand{\uvec}[0]{{\bm u}}

\begin{document}

%\preprint{APS/123-QED}

\title[]{"Hall" transport of liquid crystal solitons in Couette flow}% Force line breaks with \\

\author{Rodrigo C. V. Coelho}
\affiliation{Centro de Física Teórica e Computacional, Faculdade de Ciências, Universidade de Lisboa, 1749-016 Lisboa, Portugal.}%Lines break automatically or can be forced with \\
 \affiliation{Departamento de Física, Faculdade de Ciências, Universidade de Lisboa, P-1749-016 Lisboa, Portugal.}
\email[]{rcvcoelho@fc.ul.pt}
\author{Hanqing Zhao}
\affiliation{Department of Physics and Soft Materials Research Center, University of Colorado Boulder, CO 80309, USA.}
\author{Guilherme N. C. Amaral}
\affiliation{Centro de Física Teórica e Computacional, Faculdade de Ciências, Universidade de Lisboa, 1749-016 Lisboa, Portugal.}%Lines break automatically or can be forced with \\
 \affiliation{Departamento de Física, Faculdade de Ciências, Universidade de Lisboa, P-1749-016 Lisboa, Portugal.}
\author{Ivan I. Smalyukh}
\email{ivan.smalyukh@colorado.edu}
\affiliation{Department of Physics and Soft Materials Research Center, University of Colorado Boulder, CO 80309, USA.}
%\todo[size=\tiny]{I edit the affliation, please check}
\affiliation{Department of Electrical, Computer, and Energy Engineering and Materials Science and Engineering Program, University of Colorado, Boulder, CO 80309.}
\affiliation{Renewable and Sustainable Energy Institute, National Renewable Energy Laboratory and University of Colorado, Boulder, CO 80309, USA.}
\affiliation{International Institute for Sustainability with Knotted Chiral Meta Matter, Hiroshima University, Higashihiroshima 739-8511, Japan.}
\author{Margarida M. Telo da Gama}% 
  \affiliation{Centro de Física Teórica e Computacional, Faculdade de Ciências, Universidade de Lisboa, 1749-016 Lisboa, Portugal.}%Lines break automatically or can be forced with \\
 \affiliation{Departamento de Física, Faculdade de Ciências,
Universidade de Lisboa, P-1749-016 Lisboa, Portugal.}
\author{Mykola Tasinkevych}
\email{mykola.tasinkevych@ntu.ac.uk}
  \affiliation{Centro de Física Teórica e Computacional, Faculdade de Ciências, Universidade de Lisboa, 1749-016 Lisboa, Portugal.}%Lines break automatically or can be forced with \\
 \affiliation{Departamento de Física, Faculdade de Ciências,
Universidade de Lisboa, P-1749-016 Lisboa, Portugal.}
\affiliation{SOFT Group, School of Science and Technology, Nottingham Trent University, Clifton Lane, Nottingham NG11~8NS, United Kingdom.}
\affiliation{International Institute for Sustainability with Knotted Chiral Meta Matter, Hiroshima University, Higashihiroshima 739-8511, Japan.}


\begin{abstract}
Topology establishes a unifying framework for a diverse range of scientific areas including particle physics, cosmology, and condensed matter physics. One of the most fascinating manifestations of topology in the context of condensed matter is the topological Hall effect, and its relative: the Skyrmion Hall effect.  Skyrmions are stable vortex-like spin configurations in certain chiral magnets, and when subject to external electric currents can drift in the transverse direction to the current. These quasi-particles are characterised by a conserved topological charge which in the Skyrmion Hall effect plays the role of electric charges in the ordinary Hall effect. Recently, it has been shown that liquid crystals endowed with chiral properties serve as an ideal testbed for the fundamental investigation of topological solitons, including their two- and three-dimensional realisations. Here, we show experimentally and numerically that three-dimensional solitons aka "torons" exhibit a Hall-like effect when driven by shear flows: the torons are deflected in the direction perpendicular to the shear plane. The experimental results are rationalised in terms of the dynamic Ericksen-Leslie equations, which predict the emergence of the transverse component of the net mass flow, the magnitude of which scales as the 3rd power of the shear rate. The perturbation analysis highlights an interplay of the viscous and chiral elastic torques as the mechanism for the emergence of net transverse currents. Numerical simulations demonstrate, however, that torons are not merely dragged by the flow but move with their own transverse speed, much larger than the average flow velocity in the transverse direction. Our findings may enable responsive microfluidic applications relying on soft topological solitons. 

%Topology establishes a unifying framework for a diverse range of scientific areas including particle physics, cosmology, and condensed matter physics. One of the most fascinating manifestation of topology in the context of condensed matter is the Skyrmion Hall effect. Skyrmion is an emergent property of a spin configuration and is characterized by conserved topological charge, which in the Skyrmion Hall effect play the role of electric charges in the classical Hall effect. It has recently been shown that chiral liquid crystals enable a large variety of topological solitons including two-dimensional skyrmions. Here, we show experimentally and numerically that three-dimensional solitons aka "torons" exhibit a Hall-like effect when driven by shear flows: the torons are deflected in the direction perpendicular to the shear plane. The experimental results are rationalised in terms of the Ericksen-Leslie model, which predicts the emergence of a net transverse flow whose magnitude scales as the 3rd power of the shear rate. Numerical simulations demonstrate, however, that torons are not merely dragged by the flow but move with their own transverse speed, much larger than the average flow velocity in the transverse direction. Our findings may enable responsive microfluidic applications relying on soft topological solitons.
 
\end{abstract}

\maketitle

%\tableofcontents

\section{Introduction}

Liquid crystals (LCs) combine fluidity and anisotropy in a single condensed phase and as such exhibit a diversity of unparalleled properties. Examples include facile response of the LC order parameter to external stimuli, which is used in advanced photonic applications and in reconfigurable on demand composite nanomaterials \cite{hess:2020,papic:2021,li:2017,Mertelj2013,Mundoor:2016,Mundoor2021}. In its simplest realisation a LC is a fluid of rod-like molecules, which under favorable conditions tend to align along a common direction known as the LC director $\nvec$. This state of matter is known as the nematic phase. It exhibits an elastic character due to the director's intrinsic tendency to remain spatially uniform, $\nvec(\rvec) = \nvec_0$, as dictated by the free energy minimisation. This intrinsic tendency to uniformity may be altered by introducing small amounts of chiral additives that give rise to helical director patterns as the ground state configurations of the system \cite{deGennes1995}. The resulting chiral nematic phase may be subsequently utilised as a soft template to transfer its chirality and periodicity into nanomaterials with interesting functionalities \cite{Zhang2022}. On the other hand, endowing LCs with chirality transforms these systems into unique testbeds for the fundamental exploration of topology. This branch of pure mathematics provides unifying concepts that link condensed matter physics with high energy physics and cosmology.

Chiral LCs enable a large variety of thermodynamically stable topological solitons, i.e. spatially localized robust distortions of the director field, which behave as quasi-particles. For instance, two-dimensional (2D) skyrmions \cite{Fukuda2011,Ackerman2014,Posnjak2016}, and three-dimensional (3D) hopfions \cite{Ackerman2017b,voinescu:2020}, torons \cite{Smalyukh2010,guo:2016}, heliknotons, \cite{tai:2019} and m\"{o}biusons \cite{Zhao2023} have been reported experimentally. Surprisingly, some of these structures are hybrids combining singular and non-singular topological defects. For instance, a toron is a hybrid with  skyrmion-like and Hopf fibrations features, which in addition is decorated by two self-compensating point defects required by  perpendicular alignment boundary conditions of the director field at confining surfaces.
%of two hedgehogs and a Hopf fibration, each characterised by its own topological invariant/charge. 
Skyrmions, hopfions and torons have  been also observed in chiral magnets \cite{Fert2013,Nagaosa:2013,Sutcliffe_2018,mueller:2020,Kent2021,Zheng2023}. The spatial extension of these solid state solitons is $\sim 10-100nm$,  to be compared with the size $\sim 10\mu m$ of their soft matter counterparts.

One of the most fascinating manifestations of topology in chiral magnets, is the emergence of effective electromagnetic fields when the non-coplanar spin configurations of skyrmions are coupled to the conduction electrons \cite{Volovik_1987,Schulz2012,Nagaosa:2013}. The skyrmion magnetization profile has a non-zero scalar spin chirality, which acts on the conduction electrons as an effective magnetic field described by the gauge field $a_{\mu}, (\mu=x,y)-$an  analog of the electromagnetic vector potential. This is the basis of the topological Hall effect whereby the electrons are deflected by the Lorentz force due to the emergent magnetic field \cite{Nagaosa:2013}. Alternatively, the electron current with density $j_\mu$ leads to the transverse (with respect to the current) deflection of skyrmions via the spin-transfer torque arising from the coupling $j_\mu a_\mu$--known as the "Skyrmion Hall effect" \cite{Litzius2017,Jiang2017}. The mechanism is most transparent when the skyrmion dynamics is considered in terms of the coarse-grained Thiele equation \cite{thiele:1973,everschor:2012}. The latter treats the skyrmion as a rigid body and contains explicitly the coupling between the skyrmion topological charge and the skyrmion velocity, which renders the transverse velocity component non-zero \cite{Litzius2017}.

In this article we pose the following question: can LC solitons undergo a similar current driven Hall-like transport, as both hard and soft systems are governed by similar continuum equations? It is well known, that LC solitons can be brought into directional motion by time dependent electric fields \cite{Ackerman2017,Sohn2018,sohn2020, Zhao2023}, but little is known on the interactions of LC skyrmions with material currents.

In what follows, we address this question using experiments, numerical simulations and perturbation analysis. The experiments reveal the transverse deflection of torons in shear flow$-$a behaviour akin to the Skyrmion Hall effect in magnets. The theoretical analysis uncovers the mechanism responsible for the transverse current. The numerical simulations shed light on the evolution of the toron structure with increasing shear rate and establish the dependence of the toron's angle of deflection on the shear rate. The simulations also show that torons are not simply advected by the currents but move much faster than the later.

\begin{figure}[htb]
\includegraphics[width=0.49\textwidth]{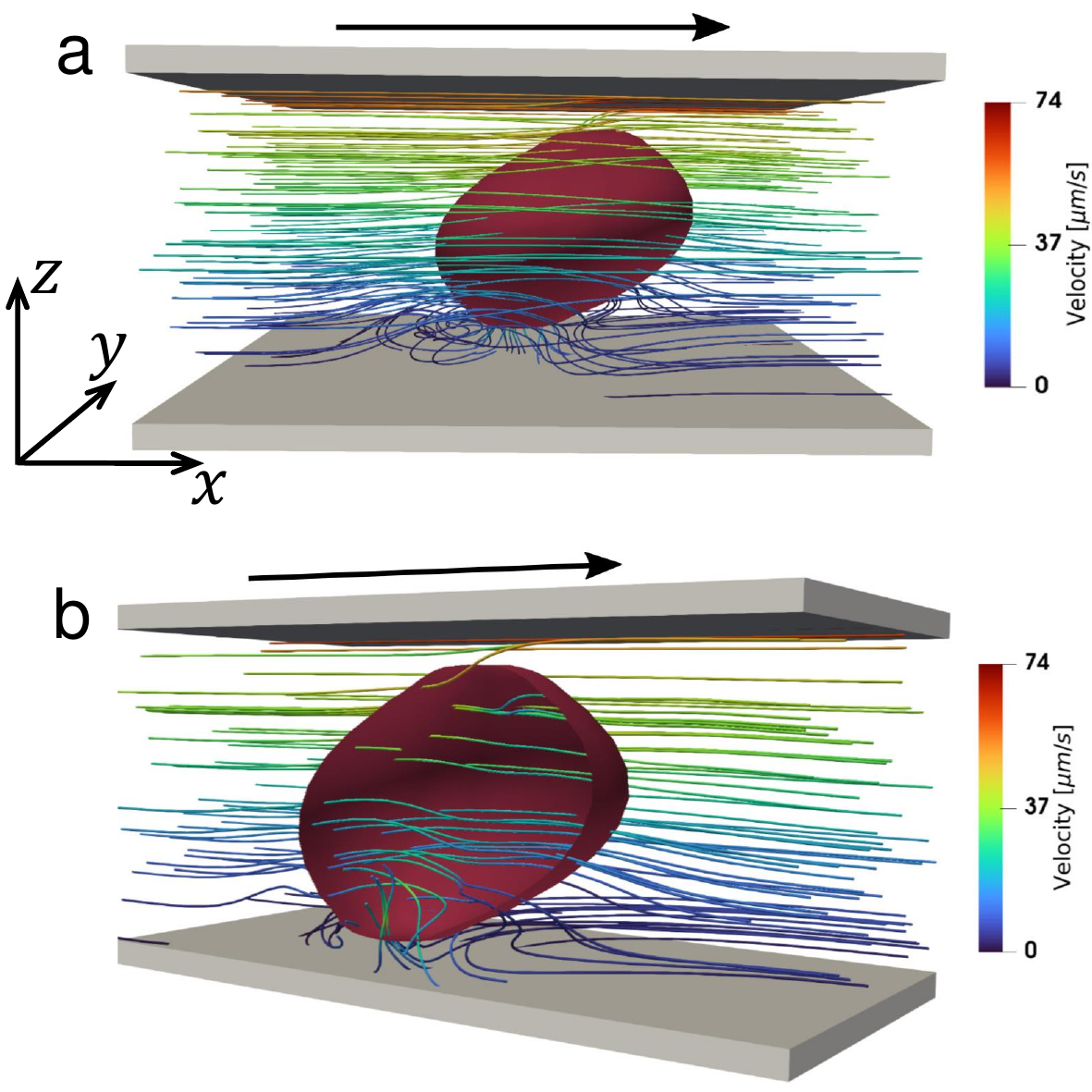}
\caption{Schematic illustration of the system geometry and representative configuration of a liquid crystal toron in Couette flow, characterised by Ericksen number $Er=4.61$, see main text for details. The liquid crystal is confined between two parallel plates. The lower plate is fixed while that on the top moves with constant speed in the direction indicated by the black arrow. The curves represent fluid streamlines which are colored to depict the magnitude of the flow velocity. The red surface in the middle represents the toron shape, corresponding to the isosurface where the $z-$component of the director field $n_z=1$ ($n_z=-1$ far from the toron). (a) Depiction of the flow field outside the toron, while in (b) the flow inside the toron is also shown. The confining parallel plates impose rigid homeotropic boundary conditions of the LC director.}
\label{figure1}
\end{figure}

\section{Results}
\label{sec:res}

We start by describing the experimental results on cross-stream migration and elongation of torons in rectilinear shear flow applied in the $x-$direction, followed by a perturbation analysis of the Ericksen-Leslie equations in powers of the shear rate. The analysis does not include torons, which are dealt with in the final subsection by numerically solving the Ericksen-Leslie equations using a hybrid method based on lattice Boltzmann and finite differences.

Figure \ref{figure1} illustrates both the system geometry and a typical result of the simulations depicting the flow streamlines and the shape of the toron. In what follows we consider rigid homeotropic alignment of the LC director field at the confining parallel plates at $z=0$ and $z=h$, and vary the shear rate by changing the speed of the upper plate, while keeping the lower plate fixed.

\subsection{Experiments}
\label{sec:exp}

\begin{figure}[htb]
\includegraphics[width=0.47\textwidth]{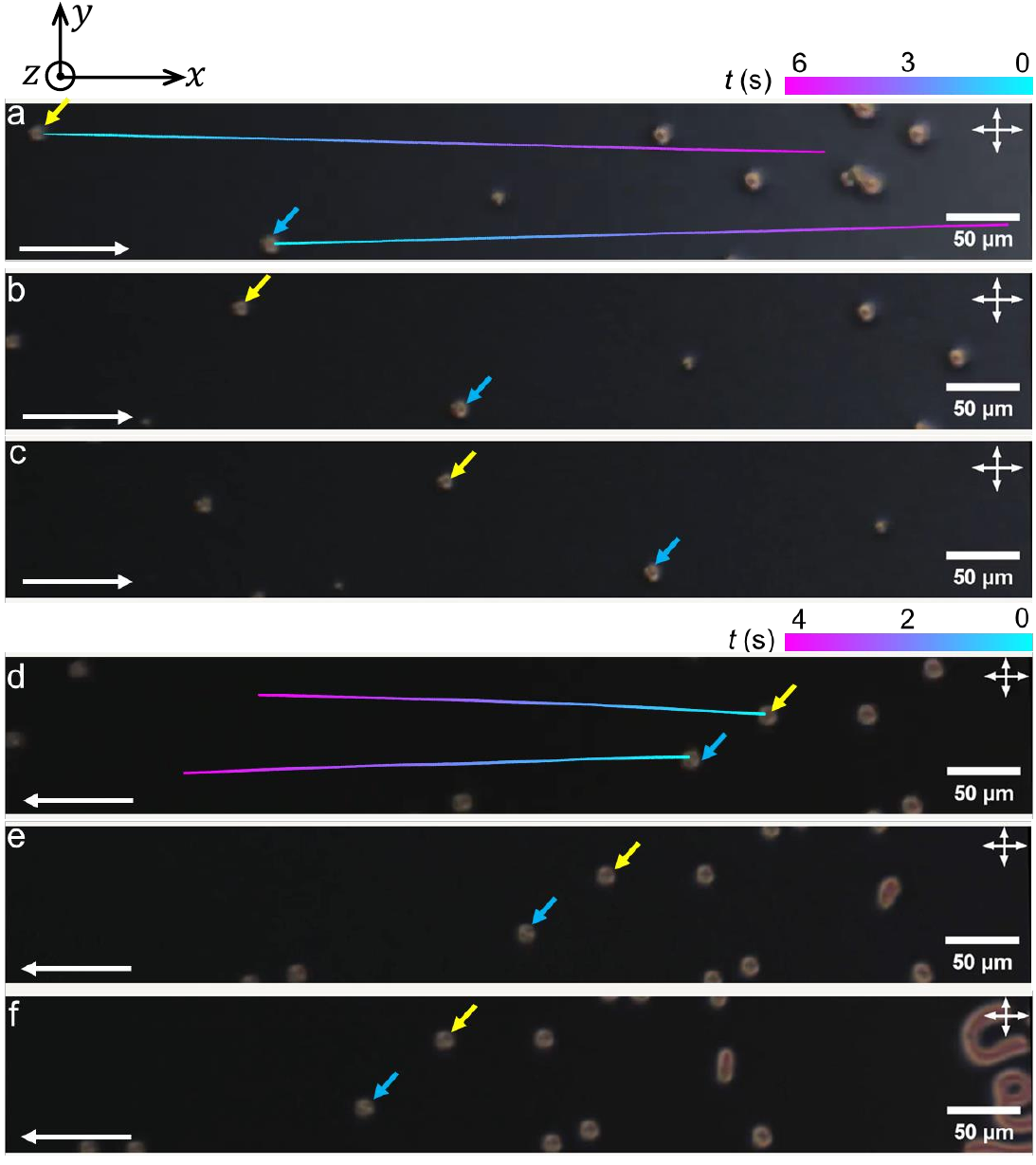}
\caption{Micrographs taken at crossed polarisers (schematically shown by two crossed white arrows at the right upper corners of the panels) demonstrating Hall-like cross-stream motion of the torons in Couette flow. The upper glass plate of the LC cell is translated horizontally relative to the lower plate. The upper plate moves from left-to-right with speed $v_p=68\,\mu m/s$, which corresponds to Ericksen number $Er =2.95$
%calculated at cholesteric pitch $P=10 \mu m$, viscosity $\mu = 28 mPa\cdot s$, and average elastic constant $K=6.46 pN$,
on (a)-(c), and from right-to-left with speed $v_p=55\,\mu m/s$, $Er=2.39$, on (d)-(f). The displacement direction of the upper plate is indicated by the horizontal white arrows at the left lower corners of the panels. The micrographs are taken at times $t=0$ (a) and (d); $t=2s$ (b) and (e); $t=4s$ (c) and (f). The toron trajectories, reconstructed from videos, are shown in (a) and (d) by colored curves, where the color encodes time according to the color bar. The torons exhibit a non-zero transverse velocity perpendicular to the shear plane $(x,z)$. The inclined yellow and blue arrows point at the torons which follow the trajectories illustrated in (a) and (d). The two selected torons drift in opposite directions along the transverse $y$ axis. The scale bar represents 50 $\mu$m. }
\label{figure1_exp}
\end{figure}

\begin{figure}[htb]
\includegraphics[width=0.47\textwidth]{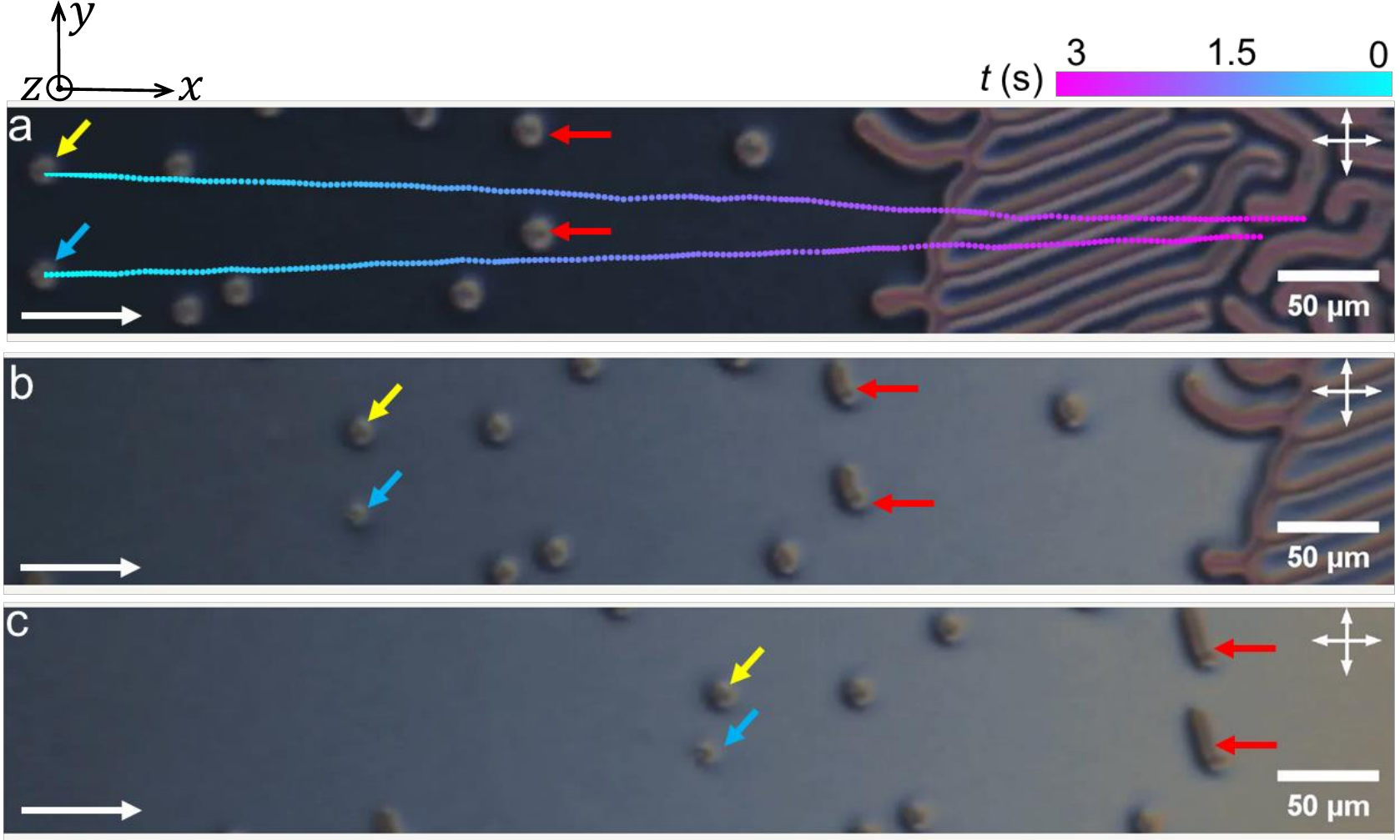}
\caption{Micrographs showing the motion of torons in Couette flow when the upper glass plate of the LC cell is translated horizontally with speed $v_p=172\,\mu m/s$, $Er = 7.46$, relative to the lower plate in the $x$ direction, shown by the horizontal white arrows at the left lower corners of the panels. The micrographs are taken at times $t=0$ (a),  $t=1s$ (b) and $t=2s$ (c). The toron trajectories are shown in (a) by the colored curves, where the color encodes time according to the color bar. The torons exhibit non-zero velocity perpendicular to the shear plane. The inclined yellow and blue arrows point at the torons which follow the trajectories in (a) and (d). The two selected torons drift in the opposite direction along the transverse $y$ axis. The red arrows show torons that elongate in the direction perpendicular to the shear plane. The scale bar represents 50 $\mu$m.}
\label{figure2_exp}
\end{figure}

%-----------------figure-------------------------------
\begin{figure}[htb]
\center
\includegraphics[width=0.95\linewidth]{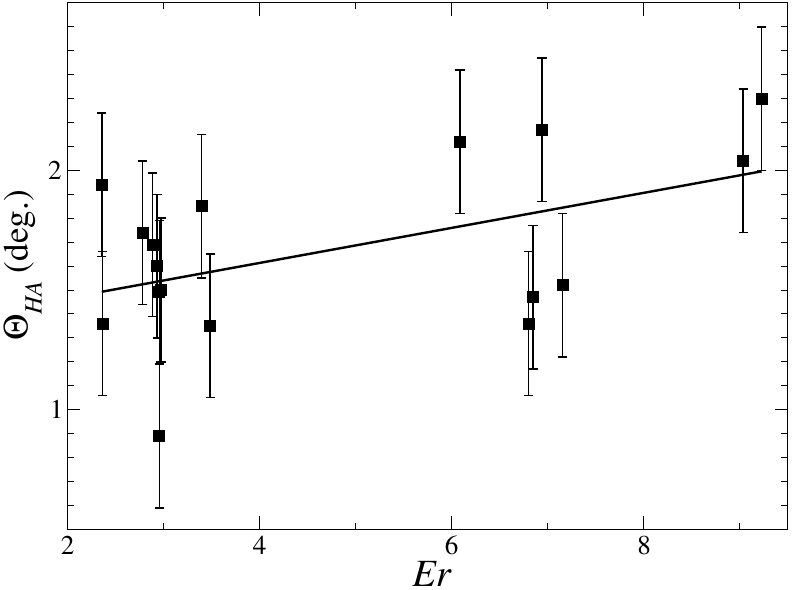}
\caption{Toron Hall angle $\Theta_{HA} \equiv \arctan(v_y/v_x)$ as a function of the Ericksen number $Er$, where $v_y, v_x$ are the $y$ and $x$ components of the toron velocity. The solid line is a linear fit to the experimental data.} 
\label{fig:Theta_vs_Er_exp}
\end{figure}
%------------------------------------------------------

%-----------------figure-------------------------------
\begin{figure}[htb]
\center
\includegraphics[width=\linewidth]{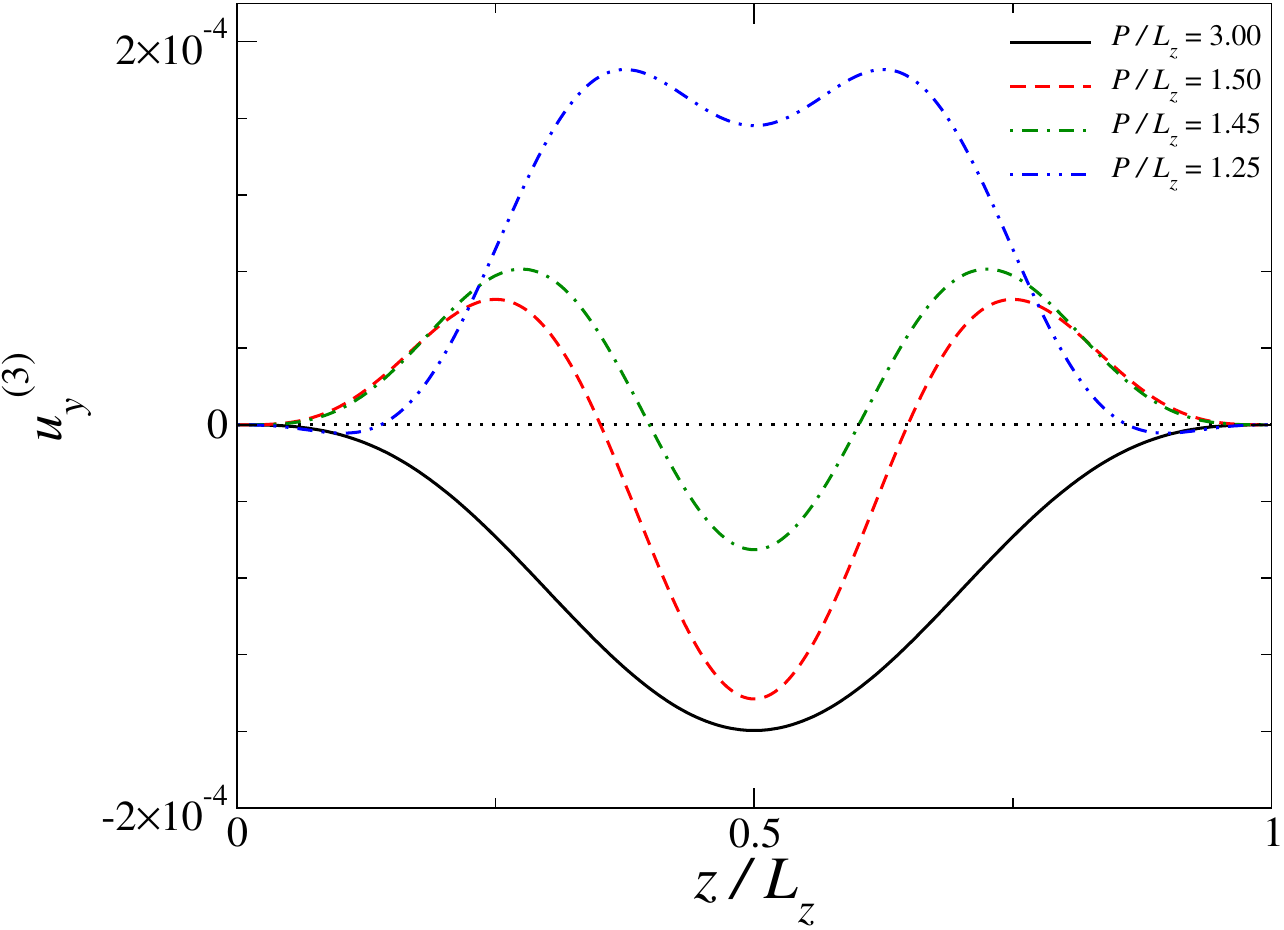}
\caption{${\cal O} (\varepsilon^3)$ dimensionless contributions $u_y^{(3)}(z)$ to the out of the shear plane component of the flow field, obtained using supplementary equation (42). $\varepsilon \equiv \frac{\alpha_4 \dot{\gamma} L_z}{q_0 K}$ is the perturbation parameter which is proportional to the Ericksen number as defined in the section on experiments. The plots are obtained at different values of the pitch to cell thickness ratio $P/L_z$. Values for the Leslie viscosities $\alpha_1,..,\alpha_6$ are provided in table \ref{tab1} in Methods. The velocity is in units of $\dot{\gamma} L_z$.} 
\label{fig_Uy3}
\end{figure}
%------------------------------------------------------

The experimental LC cell is composed by two parallel glass plates separated by freely rolling silica spheres which define the cell thickness $h$. This set up allows to impose shear on the confined LC, and in practice we fix one of the plates, and attach the other to the microscope stage, which is moved electronically in one direction with speed $v_p$. We use local laser-induced melting of the LC order to generate torons that emerge spontaneously. We introduce the Ericksen number $Er\equiv\frac{\mu v_p P}{K}$, where $P$ is the cholesteric pitch, $\mu$ is the viscosity, and $K$ is the average elastic constant.

The motion of torons in the direction perpendicular to the shear plane ($y$) is illustrated by polarizing optical micrographs (POMs) in Fig.~\ref{figure1_exp}(a)-(c), when the lower plate (attached to the microscope stage) is moving along the $x$ direction with speed $v_p=68\,\mu m/s$ ($Er=2.95$). Unexpectedly, we observe the deflection of torons both in $+y$ and $-y$ directions, see the yellow and blue arrows in the figure.  By reversing the direction of shear the toron velocities also reverse, as shown in Fig.~\ref{figure1_exp}(e)-(f). The full dynamics corresponding to  Figs.~\ref{figure1_exp}(a)-(c) and Figs.~\ref{figure1_exp}(d)-(f) is presented in supplementary videos 1 and 2, respectively.
%At the shear rate $v_p=68\,\mu m/s$, Fig.~\ref{figure1_exp}(a)-(c), we obtain for the toron Hall angle $\Theta_{HA} \equiv \arctan(v_y/v_x)\approx 1.3^{\circ}$, where $v_x, v_y$ are the two in-plane velocity components. This system is also shown in supplementary video 1.

Increasing the shear rate has pronounced effects both on the director configuration away from the torons, as evidenced by the change of the light transmission intensities in Figs.~\ref{figure2_exp}(a) through (c), and on the toron structure (see the toron structural changes indicated by the blue arrows). The toron roughly halves its lateral extension after $1s$, compare panels (a) and (b) in Fig.~\ref{figure2_exp}, and subsequently stays with this shape, see supplementary video 3. Additionally, the director distribution around the sheared torons loses the initial axial symmetry, and resembles that induced by a sufficiently strong voltage in experiments using LCs with negative dielectric anisotropy \cite{Ackerman2017}. The rear toron sides, relative to the shear flow, are dark (the director is oriented perpendicular to the plane of the image) compared to the much brighter frontal sides (the director has an in-plane component).

We also observe that some torons elongate predominantly in the $y$ direction, i.e. perpendicular to the shearing direction, as pointed by the red arrows in Fig.~\ref{figure2_exp}. A more careful analysis reveals that this effect is due to the pinning of one of the toron's hedgehogs to the closest cell plate. Indeed, as can be seen in supplementary video 3, the lower parts of these elongating torons are immobilised because of the aforementioned pinning. Despite these structural changes, most of the torons retain their original quasi-spherical shape, which is in sharp contrast to our earlier study \cite{PhysRevResearch.5.033210} where we reported strong elongation of torons in pressure-induced Poiseuille flow. In the latter the torons moved and elongated always in the flow direction.  

The toron Hall angle $\Theta_{HA} \equiv \arctan(v_y/v_x)$, where $v_x, v_y$ are the two in-plane velocity components, demonstrate a weak increasing trend with increasing  shear rate as shown in Fig.~\ref{fig:Theta_vs_Er_exp}. 
%however the reported effect is rather week taking into account the size of the error bar.
%For instance, for $v_p = 55\mu m/s$ ($Er=2.39$) depicted in Figs.~\ref{figure1_exp}(d)-(f) we measure $\Theta_{HA}\approx 2.2^{\circ}$. The full dynamics corresponding to  Figs.~\ref{figure1_exp}(d)-(f) is presented in supplementary video 2.
This behavior is in  qualitative agreement with the magnetic skyrmion Hall effect, where the skyrmion Hall angle increases linearly with the driving electric current density \cite{Litzius2017,Jiang2017}. However, as we will discuss later, this behavior is not universal and the existence of systems where $\Theta_{HA}$ decreases with the drive is not excluded.

To gain understanding on the experimental results reported above, we first consider the effects of shearing upon the uniform unwound cholesteric, i.e. the uniform director field aligned perpendicular to the confining plates. The effect of a rectilinear shear upon a uniform nematic LC, initially aligned in the shear plane perpendicular to the flow velocity, was studied extensively both experimentally and theoretically using the Ericksen-Leslie model, and interested readers are referred to this review paper \cite{Dubois-Violette1996}. In the context of the work reported here, the most relevant finding is that at small shear rates the in-plane director configuration is stable both in the flow aligning (as in the system under study) and tumbling regimes. In the latter regime the in-plane configuration becomes unstable above a threshold shear rate \cite{Dubois-Violette1996}. 

In what follows, we show, that for a cholesteric LC the $y$-component of the director $n_y$ perpendicular to the shear plane behaves as $n_y\sim\dot{\gamma}$, supplementary equation (26), where $\dot{\gamma}$ is the shear rate. Non-zero $n_y$ will in turn become a source for the transverse mass flow $u_y\sim\dot{\gamma}^3$,  supplementary equation (41).   

%These raises the following questions: Why torons move in opposite directions? Why some structures start deforming in the direction perpendicular to the primary flow? Is there a transverse secondary flows?

%\cite{fluids3030047} LB simulations of sheared cholesteric in quasi 2D geometry show the presence of the secondary flow nperpendicular to the shear plane.

\subsection{Perturbation analysis of Couette flows imposed upon unwound cholesterics}

Here, we adopt the Ericksen-Leslie model for the nematodynamics, as described in the Methods below, and construct perturbation solutions for the steady state director $\nvec(z)$ and flow $\uvec(z)$ fields as series expansions in powers of the small parameter $\varepsilon = \frac{\alpha_4 \dot{\gamma} L_z}{q_0 K}\propto Er$, where $z$ is the coordinate along the axis perpendicular to the plates, $L_z$ is the separation between the plates, $\alpha_4$ is one of the Leslie's viscosity whose value is provided in table~\ref{tab1} in "Methods", and $q_0 = 2\pi/P$, with $P$ the cholesteric pitch. 
The unperturbed zeroth order configurations are $\uvec = (0,0,0)^T$ and $\nvec = (0,0,1)^T$, and the boundary conditions read $\uvec(z=0)=(0,0,0)^T$ at the lower boundary and $\uvec(z=L_z)=(\dot{\gamma} L_z,0,0)^T$ at the upper one, and we use $\nvec(z=0,L_z) = (0,0,1)^T$ at both boundary planes. We also nondimensionalise the flow field with $\dot{\gamma} L_z$. The details of the perturbation approach are presented in supplementary material.

The first order, ${\cal O} (\varepsilon)$, contributions $u_{x,y}^{(1)}(z)$ to the velocity field are not affected by the director field, and obey the Laplace equation. With the considered boundary conditions we find $u_x^{(1)}(z) = z L_z^{-1}$, $u_y^{(1)}(z)=0$. The flow configuration gives rise to a dynamic viscous torque exerted upon $\nvec$. The $y$-component of the torque which acts on $n_x$ can be written as $\Gamma_y^{v} = (\alpha_3 n_x^2 - \alpha_2 n_z^2) \partial_z u_x= -\alpha_2\partial_z u_x +  \mathrm{h.o.t}$, where $\mathrm{h.o.t}$ stands for "higher order terms". This ${\cal O} (\varepsilon)$ component of the viscous torque drives the director away from the ground state imparting positive ($\alpha_2 < 0$) contributions to $n_x^{(1)}(z)$. By contrast the $x$-component of the viscous torque, affecting $n_y$,  is $\Gamma_x^{v} = -\alpha_3 n_x n_y \partial_z u_x={\cal O} (\varepsilon ^3) + \mathrm{h.o.t}$ and therefore does not contribute to the $1^{\mathrm st}$ order term $n_y^{(1)}(z)$. The system is driven out of the ground state solely by the chiral component of the elastic torque $\propto \left [ \nvec \cross[\nabla \cross \nvec] \right]_x = n_z \partial_z n_x = {\cal O} (\varepsilon)+ \mathrm{h.o.t}$ (see supplementary equation (21)). The $1^{\mathrm st}$ order director components $n_{x,y}^{(1)}(z)$ are displayed in supplementary Fig.~S1, for two different values of the pitch-to-cell thickness ratio. $n_{z}^{(1)}=0$ due to the director normalisation condition $(\nvec \cdot \nvec ) = 1$. This condition also dictates that $n_{z}^{(2)}(z) = -\left ( (n_x^{(1)} )^2  +  ( n_y^{(1)} ) ^2\right )/2$, which is the only non trivial $2^{\mathrm nd}$ order contribution, displayed in supplementary Fig.~S2. 

The out of the shear plane component of the flow velocity $u_y$ appears only at $3^{\mathrm rd}$ order, and it is evoked by the source term $\propto \partial_z \left (n_x^{(1)}n_y^{(1)} \right )$ in supplementary equation (42). This equation highlights the importance of $n_y^{(1)}$ in the emergence of the transverse flow. In Fig.~\ref{fig_Uy3} we plot $u_y^{(3)}(z)$ for several values of the pitch-to-cell thickness ratio, $P/L_z$. Supplementary figures S3 and S4 compares $u_y^{(3)}(z)$ with $u_x^{(3)}(z)$ and $n_y^{(3)}(z)$ with $n_x^{(3)}(z)$, respectively. Interestingly, at intermediate values of $P/L_z$, $u_y^{(3)}(z)$ exhibits a layered structure (see the red and green curves with $P/L_z = 1.5, 1.45$ in Fig.~\ref{fig_Uy3}), where the fluid occupying the central region of the cell flows in the negative $y$-direction, while the outer layers flow in the opposite $+y$-direction. The relative flow strength in the opposite directions and as a result the net mass flow in the $y$-direction can be adjusted by varying $P/L_z$. For large/small $P/L_z$ the flow predominantly proceeds in the $+y$/$-y$-direction. 

The layered structure of the transverse flow points towards one possible explanation of why some torons move in opposite directions along $y$. The spatial extension of a toron allows it to probe different flow regimes. It is known that the size of the toron \cite{Sohn2019a} is reduced with increasing $P/L_z$ and at $P/L_z=1.25$ it is roughly $0.6 L_z$. If a toron under such conditions is located closer to one of the plates, then it is dragged by the transverse flow into the $+y$-direction. By contrast, a toron stabilized in the central region may be dragged into the $-y$-direction, for adequately tuned parameters.

The results of the perturbation analysis can be summarised by a cascade of effects leading to the emergence of the transverse flow component $u_y(z)$. Recall, that this secondary flow perpendicular to the shear plane is one of the driving mechanisms of the Hall-like transport of torons. The sequence of effects is as follows: 1) Setting one of the plates in shear motion generates a flow profile $u_x(z)$ whose leading order term is $\dot{\gamma} z$. 2) The first-order contribution to the velocity field $u_x^{(1)}(z)$ produces a dynamic viscous torque, which at leading order acts only on $n_x$ generating a non-zero component of the director $n_x^{(1)}(z)\sim \dot{\gamma}$. 3) The first order contribution to the director field $n_y^{(1)}(z)$ is insensitive to the viscous torque and appears only due to the chiral part of the elastic torque $\sim \partial_z n_x^{(1)}$. 4) The second order terms in the flow field and in the director field $n_{x,y}$ are identically zero. 5) A non-zero contribution to the transverse velocity $u_y(z)$ appears only at $3^{\mathrm rd}$ order in $\dot{\gamma}$, as a consequence of a non-homogeneous term $\propto \partial_z \left (n_x^{(1)}n_y^{(1)} \right )$ in the governing Laplace equation. The key behind this mechanism is the chiral term in the Frank-Oseen free energy density, which enables momentum transfer from the $x$- to the $y$-direction. 

%\mm{ To add a comment on the oscillating $n_x$ at small $P/L_z$, and why this is not observed in the LB simulations?}

The next section complements the above analysis in two directions: i) it relaxes the assumption $\varepsilon \ll 1$; and ii) it introduces a solitonic toron structure into the unwound background LC configuration. These two generalizations render analytical advances difficult, and thus we resort to numerical solutions of the flow equations based on the lattice Boltzmann method. More details are provided in the Methods section.

\begin{figure}[htb]
\includegraphics[width=0.48\textwidth]{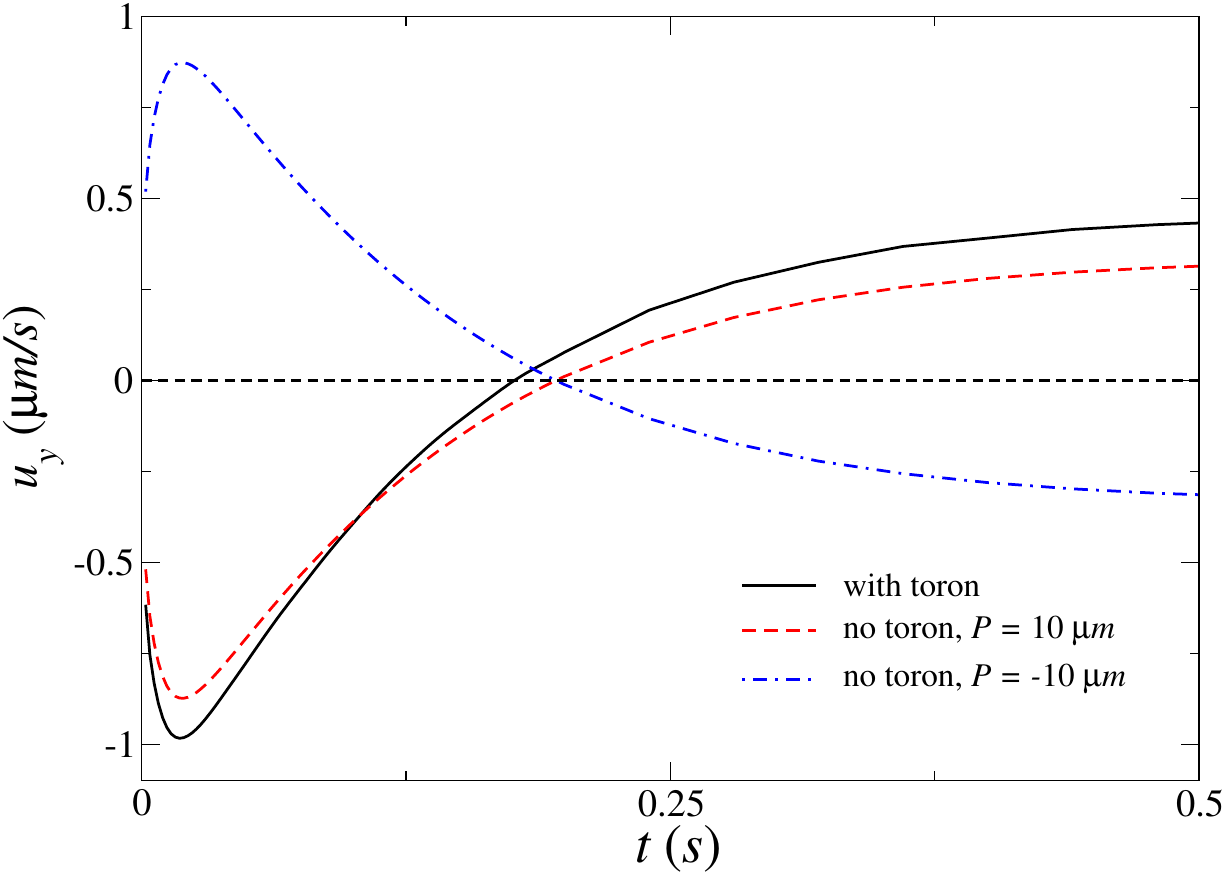}
\caption{%(a) The $y$- component $u_y$ of the steady state flow velocity as a function of the distance $z$ from the lower plate. The results are from lattice Boltzmann simulations of the Ericksen-Leslie equations \eqref{NS-eq}-\eqref{director-norm} for Ericksen number $Er \equiv \frac{\alpha_4 \langle u \rangle L_z}{2 K_{11}}\approx5.6$, where $\langle u \rangle$ is the average velocity of the fluid at steady state, $\alpha_4$ is the Leslie viscosity, $L_z$ the cell width and $K_{11}$ the splay elastic constant. (b) 
Space averaged $u_y$ as a function of time, obtained from lattice Boltzmann simulations of the Ericksen-Leslie equations \eqref{NS-eq}-\eqref{director-norm}. Results for systems with and without a toron are shown. Results for systems with opposite handedness are shown when no toron is present. The Ericksen number $Er = 3.08$.
%, where $\langle u \rangle$ is the average magnitude of the fluid velocity at the steady state, $\alpha_4$ is the Leslie viscosity, $L_z$ the cell width and $K_{11}$ the splay elastic constant.
}
\label{fig:LB_on_u_y}
\end{figure}

\begin{figure}[htb]
\includegraphics[width=0.48\textwidth]{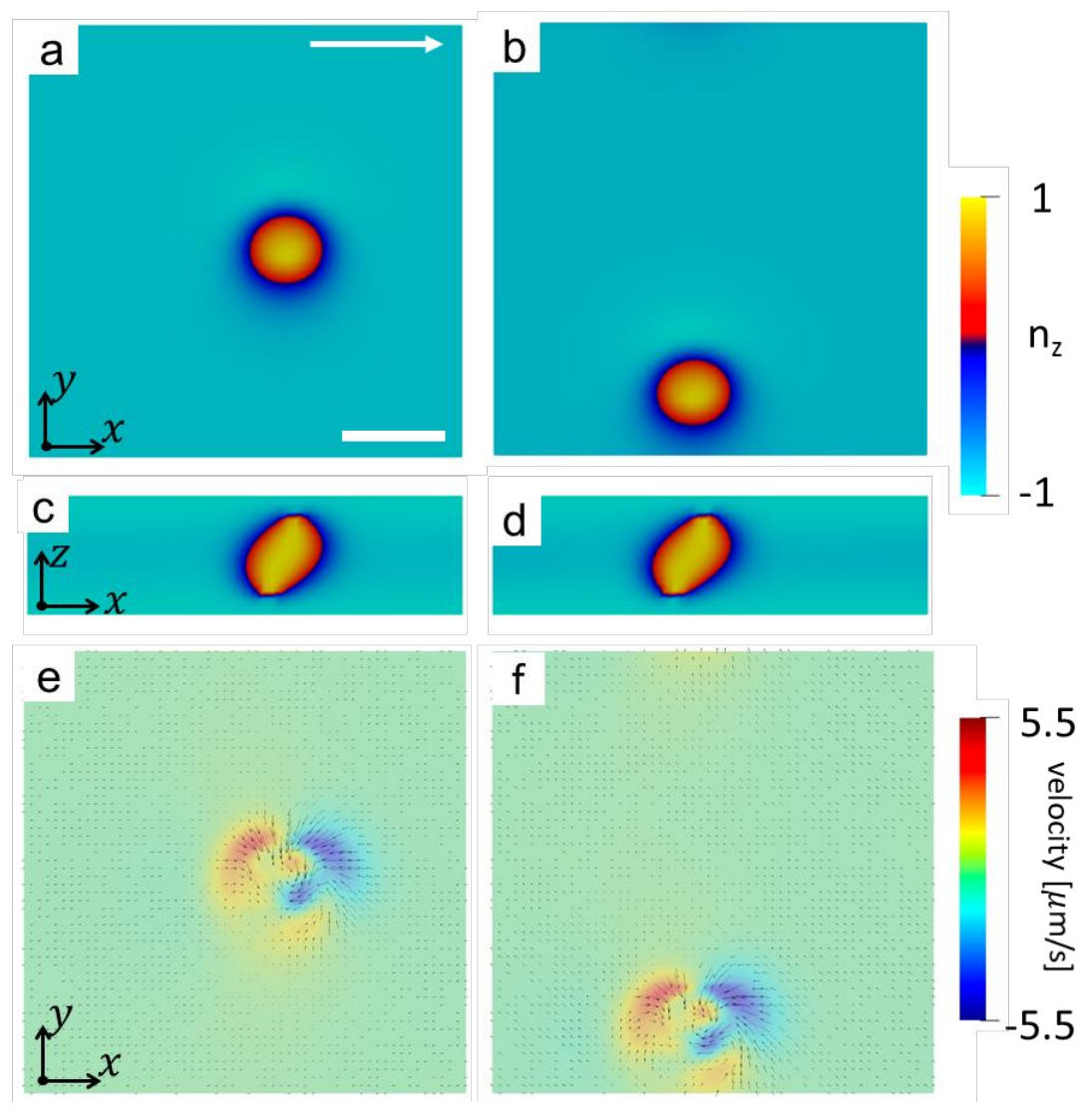}
\caption{Snapshots of simulations for $Er=4.61$ and $\Theta_{HA}=20.67^\circ$ at two different times: 0.2s and 2.3s respectively. (a) and (b) depict the $z$-component of the director field in the mid plane $z=L_z/2$. (c) and (d) are the director fields in the plane $y=y_{TCM}(t)$, where $y_{TCM}(t)$ is the instantaneous $y$-coordinate of the toron's center of mass. (e) and (f) represent the relative velocity field (local velocity minus the velocity far from the toron in the plane $z=L_z/2$). The arrows indicate the direction of the velocity field. The white arrow indicates the direction of the moving plate and the white scale bar represents 10 $\mu m$.}
\label{fig:LB_steady_skyrmion}
\end{figure}

\begin{figure}[htb]
\includegraphics[width=0.85\linewidth]{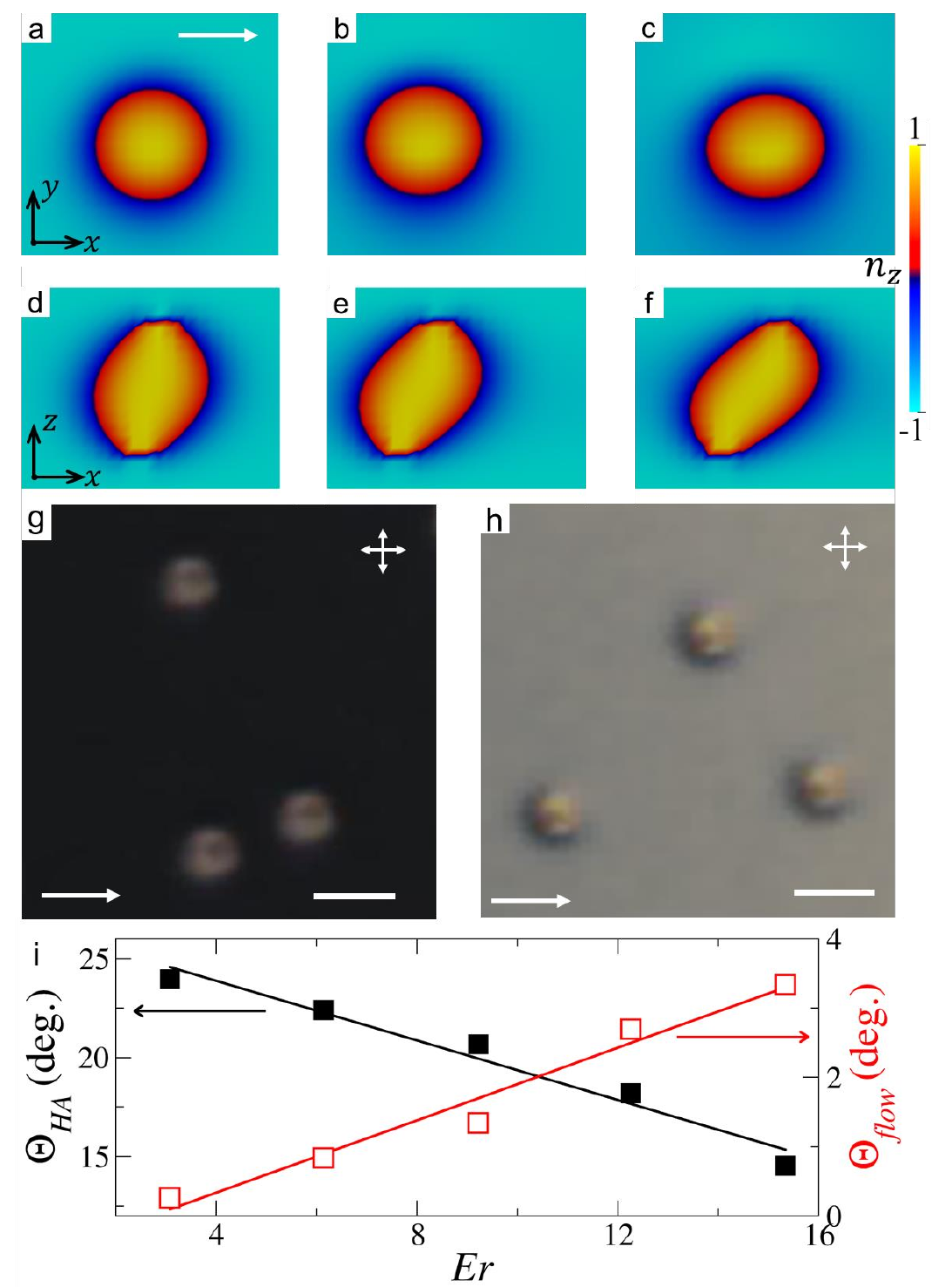}
\caption{ Torons flowing under different plate speeds. (a), (b) and (c) depict the steady state at $t=2$s in the mid plane for the Ericksen number $Er=3.08$, $Er=6.14$ and $Er=9.22$.
%where $Er\equiv \frac{\alpha_4 \langle u \rangle L_z}{2 K_{11}}$ with $\langle u \rangle$ being the space average fluid velocity at the steady state. 
(d), (e) and (f) are the corresponding figures in the plane $y=y_{TCM}(t)$, where $y_{TCM}(t)$ is the $y$-component of toron's center of mass. Experiments at low ($v_p=10\,\mu$m/s, $Er=0.43$) (g) and high ($v_p=300\,\mu$m/s, $Er=13.01$) (h) speeds illustrate the changes in the toron configuration. The scale bar represents 10$\mu$m and the arrow indicates the direction of the moving plate. (i) Hall angles $\Theta_{HA}$ as a function of $Er$ obtained from the lattice Boltzmann simulations for the toron trajectory (solid square) and the angle $\Theta_{flow}=\arctan(\langle u_y \rangle /\langle u_x \rangle )$ (empty squares) quantifying the intensity of the net mass flow in the $y$-direction, $\langle u_y \rangle$ and  $\langle u_x \rangle$ are space averaged components of the steady state flow field. The white arrows in (a), (g) and (h) indicate the direction of the moving plate. }
\label{fig:LB_steady_skyrmion_vs_Er}
\end{figure}

\begin{figure}[htb]
\includegraphics[width=0.45\textwidth]{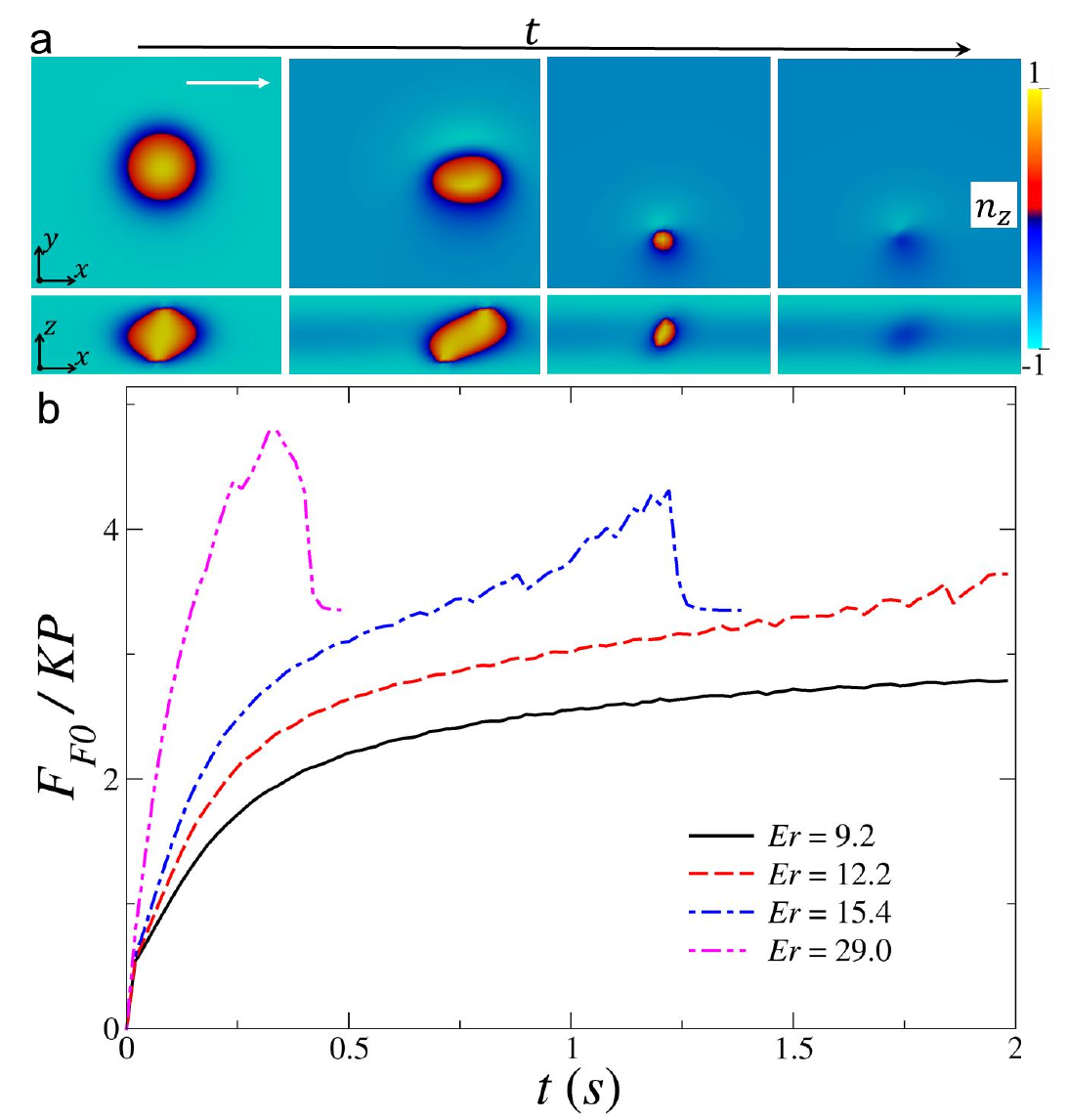}
\caption{Unstable toron at high shear rate. (a) Time evolution of the director field shown in the cross sections $z=L_z/2$ (top) and $y=y_{TCM}(t)$ (bottom), $Er=15.4$. The white arrow indicates the direction of the moving plate. (b) Time evolution of the Frank-Oseen free energy for different Ericksen numbers. }
\label{fig:LB_high_Er}
\end{figure}

\subsection{Numerical simulations}

%We define the Ericksen number as $Er\equiv \frac{\alpha_4 u_{av} L_z}{2 K_{11}}$, where $u_{av}$ is the space average fluid velocity at the steady state, $\alpha_4$ is the Leslie viscosity, $L_z$ the cell width and $K_{11}$ the splay elastic constant. 
%Here we use the same definition for the Ericksen number as in the preceding section.
Supplementary figure S5 illustrates the steady state transverse flow field $u_y(z)$ obtained from lattice Boltzmann simulations of the Erickson-Leslie equations at moderate Ericksen number $Er=5.6$, for several values of the pitch to cell width ratio, $P/L_z$. These results were obtained for systems without torons. The transverse flow velocity $u_y(z)$ agrees qualitatively with the results of the perturbation analysis in Fig.~\ref{fig_Uy3}. The transverse net current defined as the average over $z$ of the flow velocity $\langle u_y(z,t) \rangle = L_z^{-1}{\int_0}^{L_z}u_y(z,t)dz$, shown in Fig.~\ref{fig:LB_on_u_y}. Surprisingly, $\langle u_y(z,t) \rangle$ exhibits nontrivial dynamics with a local minimum/maximum for left/right handedness at early times, and a reversal of the current direction at intermediate times $\approx 0.2 s$.

In the presence of one toron, the transverse currents are similar but have slightly enhanced amplitudes (black curve in Fig.~\ref{fig:LB_on_u_y}). This effect may be due to the finite system size. As we employed periodic boundary conditions the finite size effects lead to toron-toron interactions, implying that it may be possible to control the transverse current through the number density of flowing toron lattices.   

By contrast to pressure driven flows \cite{PhysRevResearch.5.033210}, which elongate torons and skyrmions in the flow direction, the torons in Couette flow maintain their quasi-spherical shape. Figure~\ref{fig:LB_steady_skyrmion} shows the steady state cross-sectional director, Figs.~\ref{fig:LB_steady_skyrmion}(a)-(d), and flow field configurations Figs.~\ref{fig:LB_steady_skyrmion}(e) and (f). The toron moves as a rigid body in the right-downwards direction with Hall angle $\Theta_{HA}=20.67^\circ$. The relative flow velocity, in the plane $z=L_z/2$, exhibits an interesting pumping structure: the toron pulls in fluid from its front and rear and expels it along the vertical direction (normal to the viewing plane). The near-toron flows are also characterised by a complex pattern of vorticity, shown in supplementary Fig.~S6.  

By increasing the shear rate, the toron maintains its circular shape in the plane $z=L_z/2$, Fig.~\ref{fig:LB_steady_skyrmion_vs_Er}(a)-(c), but tilts in the direction of the shear as illustrated by the director profiles in planes  $y=y_{TCM}(t)$, Fig.~\ref{fig:LB_steady_skyrmion_vs_Er}(d)-(f). These findings are supported by the experiments, shown in Fig.~\ref{fig:LB_steady_skyrmion_vs_Er}(g) and (h).
%However, the real toron remains stable over a much broader range of $Er$. 
The polarizing optical micrograph in Fig.~\ref{fig:LB_steady_skyrmion_vs_Er}(h) is taken at $Er \approx 13$, and the corresponding director in the mid plane of the cell ($z=L_z/2$) tilts by $~45^{\circ}$ away from the $z$-axis as shown by the brightening of the image upon increasing $Er$ from Fig.~\ref{fig:LB_steady_skyrmion_vs_Er}(g) to (h). We also observe similar director tilt in the simulations.
%The large $Er$ director configurations resemble those obtained at moderate electric fields \cite{Ackerman2017}. 

The simulations predict decreasing Hall angles $\Theta_{HA}$ with increasing $Er$ with a rather steep gradient, Fig.~\ref{fig:LB_steady_skyrmion_vs_Er}(i) and supplementary figure S7, by contrast to the experimental results in Fig.~\ref{fig:Theta_vs_Er_exp}, which show a small positive slope of $\Theta_{HA}(Er)$. We can pinpoint at least two factors potentially influencing the toron's dynamics which have been neglected in the simulations. Firstly, the experimental cell is open at its lateral sides, which corresponds to free boundary conditions at $y=\pm L_y /2$ in the simulations. This would create complex recirculating flow patterns due to the lateral flow confinement. We have adopted periodic boundary conditions in the $y$-direction, which avoid or rather do not describe those flows. Secondly, inevitable surface imperfections acting as pinning sites could influence the motion of torons. For instance, it is known that quenched disorder creates a drive-dependent magnetic skyrmion Hall effect \cite{Reichhardt_2016}. Experiments reported in \cite{Jiang2017,Litzius2017} also show increasing $\Theta_{HA}(j)$ for certain magnetic skyrmions as a function of the current magnitude $j$. The precise form of $\Theta_{HA}(j)$, however is expected to depend on details of the specific experimental realisation, as is revealed by the theoretical analysis based on a refined version of the Thiele equation \cite{Litzius2017}. The analysis does not exclude decreasing $\Theta_{HA}(j)$.  
Surprisingly, the transverse speed acquired by torons in a steady state is much larger than the space averaged fluid velocity $\langle u_y \rangle$ in the $y$-direction. Fig.~\ref{fig:LB_steady_skyrmion_vs_Er}(i) show the angle $\Theta_{flow}=\arctan(\langle u_y \rangle /\langle u_x \rangle )$ quantifying the deflection of the average streamline in the $y$-direction, showing that $\Theta_{flow}$ is roughly 5 to 25 times smaller than  $\Theta_{HA}$. This demonstrates that torons are not merely dragged by the mass flow, but move independently on it with their own speed. Additionally, the order of magnitudes and the upward trend  of $\Theta_{flow}$ are in quantitative agreement with the experimental results for $\Theta_{HA}$ in Fig.~\ref{fig:Theta_vs_Er_exp}.

The simulated torons are not stable for $Er\gtrapprox 16$ and evolve into uniform in-plane configurations, shown in Fig.~\ref{fig:LB_high_Er}(a). The onset of the disintegration of torons is marked by an abrupt downward jump of the Frank-Oseen free energy $F_{FO}(t)$ shown in Fig.~\ref{fig:LB_high_Er}(b). 
%Figure \ref{fig:LB_steady_skyrmion_vs_Er} the toron shape is robust within a range of $\Er$, only small deformations. In experiments the toron is stable in a wider range of the Ericksen numbers. Such that even the director is strongly tilted by the flow and the toron becomes similar to the ones in moderate electric fields.
Figure \ref{fig:LB_high_Er} also suggests that there is a threshold Ericksen number $Er$ required to destabilize the torons. 

\section{Conclusions and outlook}
\label{conclusion-sec}

We found particularly intriguing dynamics of LC torons when subjected to shear driven flows. Torons undergo Hall-like drifts 
in a direction perpendicular to the shear plane. Rather unexpectedly, we found that some torons with apparently identical structure were deflected in opposite transverse directions. We rationalised these findings in terms of the dynamic Ericksen-Leslie equations. A perturbation analysis for small shear rates showed that the chiral term in the elastic torque is the key feature responsible for the transfer of linear momentum into the transverse direction with the associated transverse mass flow. The intensity of the transverse flow was found to scale as the third power of the shear rate at low shear rates. The lattice Boltzmann simulations showed that the steady state toron's trajectory deflects in the transverse direction much stronger that the average streamline of the flow field, with $\Theta_{HA} \approx 5,..,25 \Theta_{flow}$, depending on $Er$. This means that torons do not behave as passive tracer particles, but due to their intrinsic structure acquire extra speed exceeding the intensity of net mass currents in the $y$-direction. In addition, lattice Boltzmann simulations resolved the structure of torons in shear flow and yielded results for the toron Hall angle, over a wide range of shear rates. We found that $\Theta_{HA}$  decreases with the shear rate in contrast to the behavior found in experiments where $\Theta_{HA}$ exhibited a weak upward trend. The later behavior is similar quantitatively to that obtained for $\Theta_{flow}(Er)$. There could be several reasons for the discrepancy, such as: the effect of the toron's pinning on imperfections of the glass plates, or the influence of the open lateral boundaries in the direction perpendicular to the shear flow. These factors have been ignored in our simulations.  

We comment that for two particular realisations of magnetic skyrmions, the increase of the Hall angle with the electric current has also been reported in recent experiments \cite{Jiang2017,Litzius2017}. A detailed analysis of the magnetic skyrmion dynamics in terms of a generalised Thiele equation, which goes beyond the rigid body approximation and allows for skyrmion deformations in response to the driving current \cite{Litzius2017},  does not exclude the opposite behavior of $\Theta_{HA}$. Indeed, $\Theta_{HA}$ was found to depend sensitively  on several phenomenological parameters of the experimental system at hand.  

The results reported here are also in sharp contrast with the behavior of torons in pressure-induced Poiseuille flows where the torons are dragged along the flow and are stretched in the flow direction at sufficiently high flow rates \cite{PhysRevResearch.5.033210}. The distinct behaviors of torons subject to different material flows emphasise the complexity of the interactions of soft solitons with matter currents, contrasting with the solid-state realisations where electron currents can be assumed to be constant. One possible extension of the present study is the investigation of similar effects for other types of LC solitonic structures, such as hopfions, heliknotons,  m\"{o}biusons, or their macromolecule-like assemblies and crystals.

\section{Methods and Materials}
\label{method-sec}
\subsection{Experiments}

Chiral LCs are prepared by mixing 4-Cyano-4'-pentylbiphenyl (5CB, EM Chemicals) with a left-handed chiral additive, cholesterol pelargonate (Sigma-Aldrich), whose helical twisting power $h_{htp}=6.25 \mu m^{-1}$. The cholesteric pitch of the ensuing chiral LCs $P=C_{dopant}/h_{htp}$, where $C_{dopant}$ is the weight fraction of the added chiral dopant.

The sample cells are assembled from indium-tin-oxide (ITO)-coated glass slides treated with polyimide SE5661 (Nissan Chemicals) to obtain strong perpendicular (homeotropic) boundary conditions. The polyimide is applied to the surfaces by spin-coating at $2700 rpm$ for $30 s$ followed by baking (5 min at 90 $^\circ$C and then 1 h at 180$^\circ$C).
 
 The LC cell gap thickness is set by silica spheres as spacers between two surfaces to be 10$\mu m$ with the cell gap to pitch ratio $L_z/P = 1$. In this study the spacers are not bonded to the glass surfaces by glue, allowing the two plates to slide freely one with respect to the other. In experiments, we fix one glass plate, and attach the other one to the microscope stage (slide holder) that can be controlled electronically and moved in one direction with the speed $v_p$.
 We used the following values for the model parameters to obtain  the Ericksen number $Er = \frac{\mu v_p P}{K}$ reported in the manuscript: the cholesteric pitch $P=10 \mu m$, viscosity $\mu = 28 mPa\cdot s$, and the average elastic constant $K=6.46 pN$.

We utilize a ytterbium-doped fibre laser (YLR-10-1064, IPG Photonics, operating at 1064$nm$) to generate torons, by melting the LC locally with the laser power around $30mW$. When we switch off the laser tweezers, the torons are spontaneously generated as the LC quenches back. (Torons are three dimensional topological solitons where the skyrmion tube is embedded between the two substrates along the perpendicular axis, terminated by two points.) 

Polarizing optical microscopy images are obtained with a multi-modal imaging setup built around an IX-81 Olympus inverted microscope and charge-coupled device cameras (Grasshopper, Point Grey Research). Olympus objectives 20x with numerical aperture NA=0.4 are used. The speed and trajectories of the torons are analyzed by using ImageJ (freeware from NIH)

\subsection{Lattice Boltzmann simulation of the Ericsen-Leslie model}

 The dynamics of the LC director field is often described using the model proposed by Ericksen and Leslie~\cite{Ericksen1962, doi:10.1098/rspa.1968.0195, stewart2019static}. This model comprises two fundamental equations: one governing material flow and the other governing the director field. These equations are particularly effective in describing the behavior of LCs deep in the nematic or cholesteric phases.

For the velocity field, we use the Navier-Stokes equation and the continuity equation:
\begin{eqnarray}
 &&\rho \partial_t u_\alpha + \rho u_\beta \partial_\beta u_\alpha = \partial_\beta \left[ - \Pi \delta_{\alpha\beta} + \sigma_{\alpha\beta}^{v}   + \sigma_{\alpha\beta}^{e}   \right] \label{NS-eq}\\
 && \partial_\alpha u_\alpha = 0\label{cont-eq},
\end{eqnarray}
where the viscous stress tensor is given by:
\begin{eqnarray}
 \sigma_{\alpha\beta}^{v} = && \alpha_1 n_\alpha n_\beta n_\mu n_\rho D_{\mu\rho}+ \alpha_2 n_\beta N_\alpha + \alpha_3 n_\alpha N_\beta  \\
&&+ \alpha_4 D_{\alpha\beta} + \alpha_5 n_\beta n_\mu D_{\mu\alpha}  + \alpha_6 n_\alpha n_\mu D_{\mu\beta} .
\end{eqnarray}
Here $\rho$ is the fluid density, $\Pi$ is a scalar Lagrange multiplier which arises from the assumed incompressibility \eqref{cont-eq} and has the meaning of hydrostatic pressure, $\uvec$  the fluid velocity, $\nvec$ the director field, specifying the direction of preferential alignment of the LC molecules, and $\alpha_n$'s the Leslie viscosities. The kinematic transport, which accounts for the effect of the macroscopic flow field on the microscopic structure, is given by:
\begin{eqnarray}
 N_\beta = \partial_t n_\beta + u_\gamma \partial_\gamma n_\beta - W_{\beta \gamma} n_\gamma
\end{eqnarray}
while the shear rate and vorticity tensors are:
\begin{eqnarray}
 D_{\alpha\mu} = \frac{1}{2}\left (  \partial_\alpha u_\mu + \partial_\mu u_\alpha \right), \: W_{\alpha\mu} = \frac{1}{2}\left ( \partial_\mu u_\alpha - \partial_\alpha u_\mu \right).
\end{eqnarray}
The elastic stress tensor is given by:
\begin{eqnarray}
 \sigma_{\alpha\beta}^{e} = -\partial_\alpha n_\gamma \frac{\delta \mathcal{F}}{\delta (\partial_\beta n_\gamma)},
\end{eqnarray}
where $\mathcal{F}$ is the Frank-Oseen free energy:
\begin{eqnarray}
 \mathcal{F} =&& \int dV \biggl ( \frac{K_{11}}{2} (\nabla \cdot \nvec)^2 + \frac{K_{22}}{2} \left ( \nvec\cdot [\nabla \times \nvec] + q_0 \right)^2 \\ && +  \frac{K_{33}}{2}[ \nvec\times [\nabla \times\nvec]]^2 \biggr) .
 \label{free-energy-eq}
\end{eqnarray}
$K_{11}$, $K_{22}$, $K_{33}$ are the Frank elastic constants, and $q_0=2\pi/P$, with $P$ the cholesteric pitch. 
The second set of equations describes the evolution of the director field:
\begin{align}
  & -\partial_t n_\mu + \frac{1}{\gamma_1} h_\mu - \nonumber \\ 
  & \frac{\gamma_2}{\gamma_1} n_\alpha D_{\alpha\mu} - u_\gamma \partial_\gamma n_\mu + W_{\mu\gamma}n_\gamma = \Lambda n_\mu,
    \label{director-time-eq} \\
    &\nvec \cdot \nvec = 1,
    \label{director-norm}
\end{align}
where $\Lambda$ is a scalar Lagrange multiplier arising from the constraint \eqref{director-norm}, $\gamma_1=\alpha_3-\alpha_2$ is the rotational (twist) viscosity determining the rate of relaxation of the director, $\gamma_2 =\alpha_3+\alpha_2$ is known as a torsion coefficient characterising the contribution to the viscous torque arising from gradients of the velocity field. The ratio $\gamma_2/\gamma_1$ is known as the aligning parameter, where $\vert \gamma_2/\gamma_1 \vert>1$ represents flow aligning systems and $\vert \gamma_2/\gamma_2 \vert<1$ flow tumbling ones. Finally, the molecular field reads
%where $\lambda=(\alpha_3+\alpha_2)/(\alpha_3-\alpha_2)$ is the aligning parameter, with $\vert \lambda\vert>1$ for flow aligning particles and $\vert \lambda\vert<1$ for flow tumbling ones and $\gamma=\alpha_3-\alpha_2$ is the rotational viscosity. Finally, the molecular field writes: 
\begin{eqnarray}
 h_\mu = -\frac{\delta \mathcal{F}}{\delta n_\mu}.
\end{eqnarray}

The simulations used a hybrid numerical method. The velocity field is solved using the lattice Boltzmann method~\cite{kruger2016lattice,succi2018lattice} with the elastic and viscous (except by the term proportional to $\alpha_4$) stress tensors being introduced as a force term. The equation for the director field, Eq.~\eqref{director-time-eq}, is solved using a predictor-corrector finite-differences algorithm. On the solid plates, we apply infinite homeotropic anchoring and no-slip boundary conditions, which is done using the bounce-back condition with imposed plate velocity~\cite{kruger2016lattice}. In simulations we set the Lagrange multiplier $\Lambda=0$, and instead normalised the director at each iteration step of the numerical integration of Eqs.~\eqref{director-norm}. 

The simulations are initialized with the liquid at rest and directors pointing mostly perpendicularly to the plates except in the proximity of the toron whose configuration is obtained by minimizing the free energy starting from the {\it Ansatz} of Ref.~\cite{Coelho_2021}. The material parameters are close to those of MBBA at 22$^\circ$C ~\cite{PhysRevE.89.032508}, except the absolute viscosity (or, equivalently, $\alpha_4$), which is two times larger in the simulations in order to run in reasonable times. The code is parallelized in CUDA-C and the simulations run on GPUs.

\begin{table}
\caption{\label{tab1} Parameters used in the simulation and physical units.}
\footnotesize
\begin{tabular}{|p{0.28\linewidth}|p{0.28\linewidth}|p{0.28\linewidth}|}
\hline
symbol&sim. units & physical units\\
\hline
$\rho$&1&1088 Kg/m$^{3}$\\
\hline
$\Delta x$&1 & 0.625 $\mu$m\\
\hline
$\Delta t$&1 & 2$\times10^{-9}$ s\\
\hline
$K_{11}$&$1.67 \times 10^{-7}$& $6.4\times 10^{-12}$ N \\
\hline
$K_{22}$&$7.88 \times 10^{-8}$& $3.0\times 10^{-12}$ N \\
\hline
$K_{33}$&$2.62 \times 10^{-7}$& $9.98\times 10^{-12}$ N \\
\hline
$\alpha_1$& 0.0373 & 0.0036 Pa.s\\
\hline
$\alpha_2$& -0.4496 & -0.044 Pa.s\\
\hline
$\alpha_3$& -0.0203 & -0.0020 Pa.s\\
\hline
$\alpha_4$& 0.9318 & 0.091 Pa.s\\
\hline
$\alpha_5$& 0.3084 & 0.030 Pa.s\\
\hline
$\alpha_6$& -0.1617 & -0.016 Pa.s\\
\hline
$P$ & 14 & 8.75 $\mu$m \\
\hline
$L_x$, $L_y$, $L_z$& 56, 56, 16 & 35, 35, 10 $\mu$m\\
\hline
\end{tabular}\\
\end{table}
\normalsize

\section*{Acknowledgements}

We acknowledge financial support from the Portuguese Foundation for Science and Technology (FCT) under the contracts: EXPL/FIS-MAC/0406/2021, PTDC/FIS-MAC/28146/2017 (LISBOA-01-0145-FEDER-028146), PTDC/FISMAC/5689/2020, UIDB/00618/2020, UIDP/00618/2020 and 2023.09574.CPCA.A1. I.I.S. acknowledges hospitality of the International Institute for Sustainability with Knotted Chiral Meta Matter (WPI-SKCM2) at Hiroshima University, where he was partly working on this article. The experimental research was partly supported by the U.S. Department of Energy, Office of Basic Energy Sciences, Division of Materials Sciences and Engineering, under contract DE-SC0019293 with the University of Colorado at Boulder.

%\bibliography{ref}% Produces the bibliography via BibTeX.

%Control: production of eprint (1) enabled
%

\end{document}

\title[]{Supplemental Material: "Hall" transport of liquid crystal solitons in Couette flow}

\author{Rodrigo C. V. Coelho}
\affiliation{Centro de Física Teórica e Computacional, Faculdade de Ciências, Universidade de Lisboa, 1749-016 Lisboa, Portugal.}%Lines break automatically or can be forced with \\
 \affiliation{Departamento de Física, Faculdade de Ciências, Universidade de Lisboa, P-1749-016 Lisboa, Portugal.}
\email[]{rcvcoelho@fc.ul.pt}
\author{Hanqing Zhao}
\affiliation{Department of Physics and Soft Materials Research Center, University of Colorado Boulder, CO 80309, USA.}
\author{Guilherme N. C. Amaral}
\affiliation{Centro de Física Teórica e Computacional, Faculdade de Ciências, Universidade de Lisboa, 1749-016 Lisboa, Portugal.}%Lines break automatically or can be forced with \\
 \affiliation{Departamento de Física, Faculdade de Ciências, Universidade de Lisboa, P-1749-016 Lisboa, Portugal.}
\author{Ivan I. Smalyukh}
\email{ivan.smalyukh@colorado.edu}
\affiliation{Department of Physics and Soft Materials Research Center, University of Colorado Boulder, CO 80309, USA.}
%\todo[size=\tiny]{I edit the affliation, please check}
\affiliation{Department of Electrical, Computer, and Energy Engineering and Materials Science and Engineering Program, University of Colorado, Boulder, CO 80309.}
\affiliation{Renewable and Sustainable Energy Institute, National Renewable Energy Laboratory and University of Colorado, Boulder, CO 80309, USA.}
\affiliation{International Institute for Sustainability with Knotted Chiral Meta Matter, Hiroshima University, Higashihiroshima 739-8511, Japan.}
\author{Margarida M. Telo da Gama}% 
  \affiliation{Centro de Física Teórica e Computacional, Faculdade de Ciências, Universidade de Lisboa, 1749-016 Lisboa, Portugal.}%Lines break automatically or can be forced with \\
 \affiliation{Departamento de Física, Faculdade de Ciências,
Universidade de Lisboa, P-1749-016 Lisboa, Portugal.}
\author{Mykola Tasinkevych}
\email{mykola.tasinkevych@ntu.ac.uk}
  \affiliation{Centro de Física Teórica e Computacional, Faculdade de Ciências, Universidade de Lisboa, 1749-016 Lisboa, Portugal.}%Lines break automatically or can be forced with \\
 \affiliation{Departamento de Física, Faculdade de Ciências,
Universidade de Lisboa, P-1749-016 Lisboa, Portugal.}
\affiliation{SOFT Group, School of Science and Technology, Nottingham Trent University, Clifton Lane, Nottingham NG11~8NS, United Kingdom.}
\affiliation{International Institute for Sustainability with Knotted Chiral Meta Matter, Hiroshima University, Higashihiroshima 739-8511, Japan.}

\date{\today} 

\begin{abstract}
This Supplemental Material provides additional text and figures to support the discussion in the main text. 
\end{abstract}
\maketitle  
\newpage

%%%%%%%%%%%%%%%%%%%%%%%%%%%%%%%%%%%%%%%%%%%%%%%%%%%%%%%%%%%%%%%% 

\section{Ericksen-Leslie equations for nematodynamics}

The Ericksen-Leslie model \cite{Ericksen1962, doi:10.1098/rspa.1968.0195} describes the coupled dynamics of the LC director and material flow fields. It consists of two sets of equations expressing the balance of local force and torque. For the material flow field the local force balance is the Navier Stokes equation augmented with the continuity equation for the conservation of mass:
\begin{eqnarray}
 &&\rho \partial_t u_\alpha + \rho u_\beta \partial_\beta u_\alpha = \partial_\beta \left[ -\Pi \delta_{\alpha\beta} + \sigma_{\alpha\beta}^{v}   + \sigma_{\alpha\beta}^{e}   \right] \label{NS-eq}\\
 && \partial_\alpha u_\alpha = 0\label{cont-eq},
\end{eqnarray}
where the viscous stress tensor is:
\begin{eqnarray}
 \sigma_{\alpha\beta}^{v} = && \alpha_1 n_\alpha n_\beta n_\mu n_\rho D_{\mu\rho}+ \alpha_2 n_\beta N_\alpha + \alpha_3 n_\alpha N_\beta \nonumber \\
&&+ \alpha_4 D_{\alpha\beta} + \alpha_5 n_\beta n_\mu D_{\mu\alpha}  + \alpha_6 n_\alpha n_\mu D_{\mu\beta}.
\end{eqnarray}
Here $\rho$ is the fluid number density, $\Pi$ is a scalar Lagrange multiplier which arises from the assumed incompressibility and has the meaning of hydrostatic pressure, $\bm{u}$ is the fluid velocity, $\bm{n}$ is the director field (unit vector in the direction of preferential alignment of the molecules) and $\alpha_n$'s are the Leslie viscosities of the material. The kinematic transport, which represents the effect of the macroscopic flow field on the microscopic structure, is given by:
\begin{eqnarray}
 N_\beta = \partial_t n_\beta + u_\gamma \partial_\gamma n_\beta - W_{\beta \gamma} n_\gamma
\end{eqnarray}
while the strain and vorticity tensors are, respectively:
\begin{eqnarray}
 D_{\alpha\mu} = \frac{1}{2}\left (  \partial_\alpha u_\mu + \partial_\mu u_\alpha \right), \: W_{\alpha\mu} = \frac{1}{2}\left ( \partial_\mu u_\alpha - \partial_\alpha u_\mu \right).
\end{eqnarray}
The elastic stress tensor is:
\begin{eqnarray}
 \sigma_{\alpha\beta}^{e} = -\partial_\alpha n_\gamma \frac{\delta \mathcal{F}}{\delta (\partial_\beta n_\gamma)},
\end{eqnarray}
where $\mathcal{F}$ is the Frank-Oseen elastic free energy, for which we adopt one elastic constant $K=K_{11}=K_{22}=K_{33}$
\begin{eqnarray}
 \mathcal{F} =&& \frac{K}{2}\int dV \left (   \left( \nabla \cdot \nvec\right)^2+\left( \nabla \cross \nvec\right)^2 +2q_0 \, \nvec\cdot \left [\nabla \cross \nvec \right ]    \right ),
 \label{free-energy-eq}
\end{eqnarray}
 where $q_0=2\pi/P$, with $P$ being the cholesteric pitch. 
The local torque balance governs the dynamics of the director field:
\begin{eqnarray}
 -\partial_t n_\mu + \frac{1}{\gamma_1} h_\mu - \frac{\gamma_2}{\gamma_1} n_\alpha D_{\alpha\mu} - u_\gamma \partial_\gamma n_\mu + W_{\mu\gamma}n_\gamma = \Lambda n_\mu,
 \label{director-time-eq}
\end{eqnarray}
where $\Lambda$ is a scalar Lagrange multiplier arising from the constraint $(\nvec\cdot\nvec) = 1$, $\gamma_1=\alpha_3-\alpha_2$ is the rotational (twist) viscosity determining the the rate of relaxation of the director, $\gamma_2 =\alpha_3+\alpha_2$ is known as a torsion coefficient characterising the contribution to the viscous torque arising from gradients of the shear velocity. The ratio $\gamma_2/\gamma_1$ is the aligning parameter, with $\vert \gamma_2/\gamma_1 \vert>1$ for flow aligning nematics and $\vert \gamma_2/\gamma_2 \vert<1$ for flow tumbling ones. Finally, the molecular field is: 
\begin{eqnarray}
 h_\mu = -\frac{\partial {F}}{\partial n_\mu}  + \partial_\gamma\frac{\partial F}{\partial (\partial_\gamma n_\mu)}=K\nabla^2 n_\mu - 2Kq_0 \left[ \nabla \cross \nvec\right]_\mu,
 \label{local_field}
\end{eqnarray}
where $F$ is the Frank-Oseen free energy density in Eq.~(\ref{free-energy-eq}).
${\bm h}$ has the meaning of an elastic force, and the corresponding elastic torque driving the director field towards equilibrium, global or local, is
\begin{equation}
    {\bm \Gamma^{el}} = \left[ {\bm n}\cross {\bm h} \right]
    \label{el_torque}.
\end{equation}
Additionally, the viscous force and torque are given by the following expressions
\begin{eqnarray}
    {\bm g} = -\gamma_1 {\bm N} - \gamma_2 {\bm D}\cdot {\bm n}, 
    \label{vis_force}\\
    {\bm \Gamma^{v}} = \left[{\bm n}\cross {\bm g} \right].
    \label{vis_torq}
\end{eqnarray}
    
\section{Perturbative analysis of the Ericksen-Leslie equations}

Consider a cholesteric LC confined in the $z$ direction by two parallel surfaces placed at $z=0$ and $z =L$, and unconstrained in the lateral $x$ and $y$ directions. We assume that the plate at $z=L$ moves in the $+x$ direction with the speed $\dot{\gamma} L$. We find it convenient to introduce the Ericksen number in the following form
\begin{equation}
 \varepsilon = \frac{\alpha_4 \dot{\gamma} L_z}{q_0 K}.  
 \label{small_parametre}  
\end{equation}
Below we assume $\varepsilon \ll 1$ and determine the leading order terms and next-to-leading-order corrections to the steady state solutions $\nvec(z)$ and $\uvec(z)$ to Eqs.~(\ref{NS-eq}), (\ref{cont-eq}) and (\ref{director-time-eq}). We non-dimensionalise the equations by setting $q_0^{-1}$, $\dot{\gamma}^{-1}$, $\dot{\gamma} L_z$ and $Kq_0^{2}$ as the units of length, time, velocity and pressure, respectively. In the expressions below all the Leslie viscosities are rescaled by $\alpha_4$ and to keep the notation simple, we used the same symbols $\alpha_i$ for the dimensional and dimensionless viscosity parameters.  Next, we expand  $\nvec(z)$, $\Lambda$ and $\uvec(z)$ in powers of $\varepsilon$ as follows
\begin{eqnarray}
    n_x(z) = \varepsilon n_x^{(1)}(z) + \varepsilon^2 n_x^{(2)}(z)  + \varepsilon^3 n_x^{(3)}(z) + {\cal O}(\varepsilon^4), 
    \label{nx_expantion} \\
     n_y(z) = \varepsilon n_y^{(1)}(z) + \varepsilon^2 n_y^{(2)}(z)  + \varepsilon^3 n_y^{(3)}(z) + {\cal O}(\varepsilon^4), 
     \label{ny_expantion} \\
     n_z(z) = 1 + \varepsilon^2 n_z^{(2)}(z)  + \varepsilon n_z^{(3)}(z) + {\cal O}(\varepsilon^4),
     \label{nz_expantion} \\
     \Lambda(z) = \varepsilon \Lambda^{(1)}(z) + \varepsilon^2 \Lambda^{(2)}(z)  + \varepsilon^3 \Lambda^{(3)}(z) + {\cal O}(\varepsilon^4), 
     \label{LAMBDA} 
\end{eqnarray}
and the velocity
\begin{eqnarray}
    u_x(z) = u_x^{(1)}(z) + \varepsilon u_x^{(2)}(z)  + \varepsilon^2 u_x^{(3)}(z) + {\cal O}(\varepsilon^3), 
    \label{ux_expantion} \\
     u_y(z) = u_y^{(1)}(z) + \varepsilon u_y^{(2)}(z)  + \varepsilon^2 u_y^{(3)}(z) + {\cal O}(\varepsilon^3). 
     \label{uy_expantion}
     \end{eqnarray}
The selected velocity scale $\sim\varepsilon$ dictates the expansion for the flow field $\uvec(z)$ as in equations \eqref{ux_expantion} and \eqref{uy_expantion}. Recall that the fluid velocity terms in Eqs.~(\ref{NS-eq}) and (\ref{director-time-eq}) are multiplied by $\varepsilon$ after rewriting the equations in dimensionless form. $n_z^{(1)}(z)=0$ due to the condition $(\nvec\cdot\nvec) = 1$, and the incompressibility condition \eqref{cont-eq} renders $u_z=0$. The only non-zero boundary condition is $u_x^{1}(z=L_z) = 1$, with all the other terms in \eqref{nx_expantion}-\eqref{uy_expantion} set to zero at both bounding surfaces.

As we show below, the second order terms $u_\mu^{(2)}(z)=0, n_\mu^{(2)}(z)=0$ with $\mu=x,y$, therefore the first nontrivial corrections to the flow field are ${\cal O}(\varepsilon^3)$. We stress, that $n_z^{(2)}(z)\neq 0$, but this component is not of interest here. We also do not discuss the behaviour of the hydrostatic pressure $P$.
Finally, we demand that the normalization condition $(\nvec\cdot\nvec) = 1$ is satisfied at each order of $\varepsilon$. 

\begin{figure}[th]
\center
\includegraphics[width=0.49\linewidth]{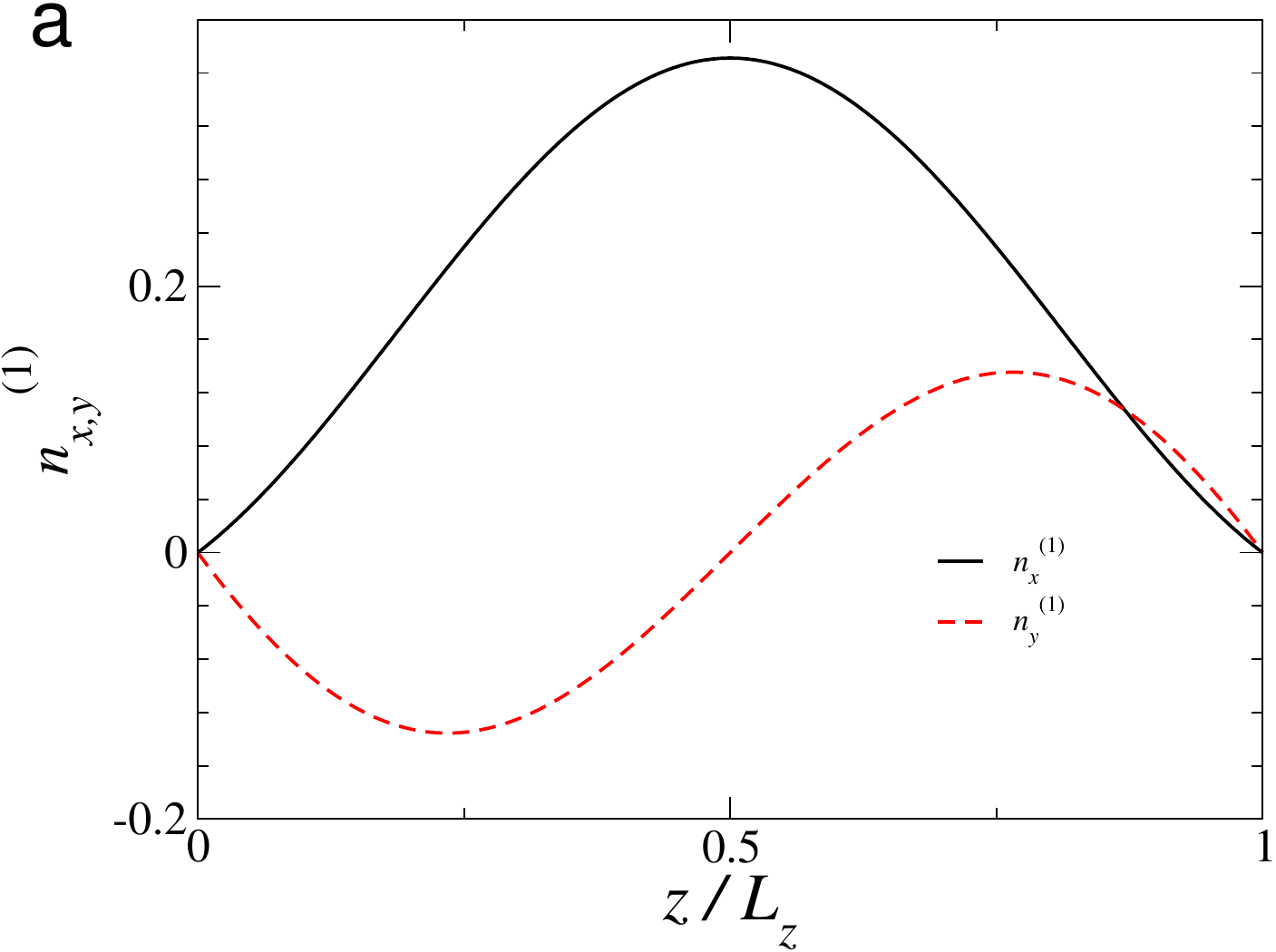}
\includegraphics[width=0.49\linewidth]{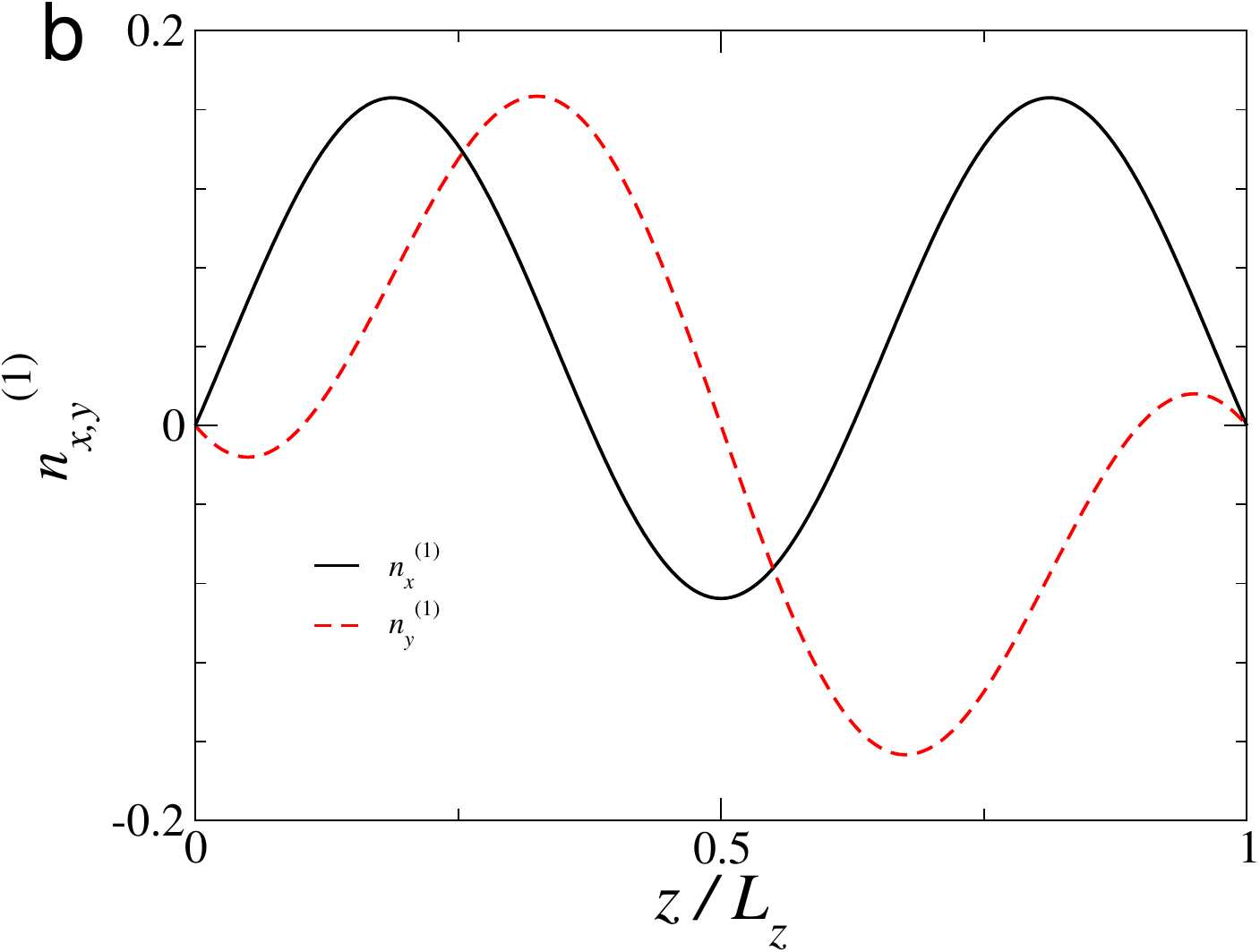}
\caption{${\cal O} (\varepsilon)$ contributions $n_x^{(1)}, n_y^{(1)}$ to the director distortions as functions of the distance $z$ to the lower surface. $n_x^{(1)}(z)$ and $n_y^{(1)}(z)$ are obtained as the solutions to \eqref{1storder_nx}-\eqref{1storder_uy}. The curves are computed at (a) $P/L_z = 2.5$, and (b) $ P/L_z = 1.25$, where $P$ is the cholesteric pitch and $L_z$ is the thickness of the LC cell.} 
\label{SI-fig_nxy1}
\end{figure}

At first order in $\varepsilon$, we find 
\begin{eqnarray}
\partial_z^2 n_x^{(1)} + 2\partial_z n_y^{(1)} - \alpha_2\partial_z u_x^{(1)} = 0,
\label{1storder_nx} \\
\partial_z^2 n_y^{(1)} - 2\partial_z n_x^{(1)} - \alpha_2 \partial_z u_y^{(1)} = 0, 
\label{1storder_ny} \\
\Lambda^{(1)} = 0, 
\label{1storder_nz} \\
\partial_z^2 u_x^{(1)} = 0,
\label{1storder_ux} \\
\partial_z^2 u_y^{(1)} = 0,
  \label{1storder_uy}  
\end{eqnarray}
where $\partial_z$ denotes a differentiation with respect to $z$. 
Equation \eqref{1storder_nx} shows that the velocity gradient $\partial_z u_x^{(1)}$ acts as a source for the $n_x^{(1)}(z)$ distortions, which in turn, Eq.~\eqref{1storder_ny}, drive the out of plane twist distortions of the director quantified by $n_y^{(1)}(z)$. This effect is absent for a pure nematic LC, where the director would stay in the shear plane. The solutions to \eqref{1storder_nx}-\eqref{1storder_uy} are
\begin{eqnarray}
n_x^{(1)}(z) = -\frac{\alpha_2}{2}\csc[L_z]\sin[L_z-z]\sin[z],
\label{1storder_solution_nx} \\
n_y^{(1)}(z) = \frac{\alpha_2}{4}\left (\frac{2z}{L_z} -1 + \csc[L_z]\sin[L_z-2z] \right ),
\label{1storder_solution_ny} \\
u_x^{(1)}(z) = \frac{z}{L_z},
\label{1storder_solution_ux} \\
u_y^{(1)}(z) = 0.
\end{eqnarray}
$n_x^{(1)}(z)$ and $n_y^{(1)}(z)$ are plotted in Fig.~\ref{SI-fig_nxy1} for two different values of the ratio pitch over the cell thickness $P/L_z$.

\begin{figure}[th]
\center
\includegraphics[width=0.65\linewidth]{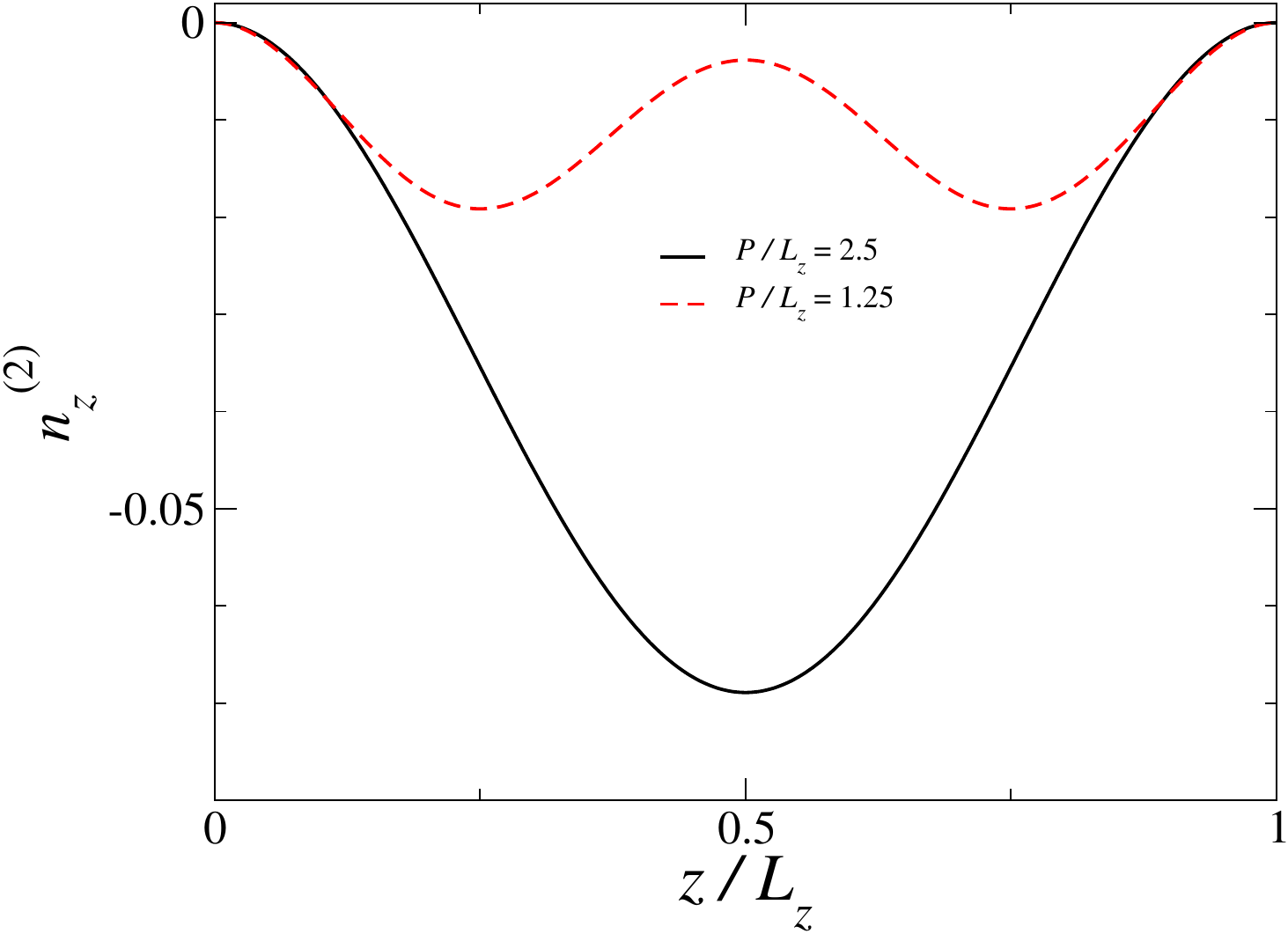}
\caption{${\cal O} (\varepsilon^2)$ contribution $n_z^{(2)}$ to the director distortions as a function of the distance $z$ from the lower surface. $n_z^{(2)}(z)$ is the solution to \eqref{2storder_nz}. The curves are computed at different values of $P / L_z$ as indicated in the legend. } 
\label{SI-fig_nz2}
\end{figure}

The second order equations are

\begin{eqnarray}
\partial_z^2 n_x^{(2)} + 2\partial_z n_y^{(2)} - \alpha_2\partial_z u_x^{(2)} = \gamma_1\Lambda^{(1)}n_x^{(1)},
\label{2storder_nx} \\
\partial_z^2 n_y^{(2)} - 2\partial_z n_x^{(2)} - \alpha_2\partial_z u_y^{(2)} = \gamma_1\Lambda^{(1)}n_y^{(1)}, 
\label{2storder_ny} \\
\partial_z^2 n_z^{(2)} - \alpha_3 L_z^{-1} n_x^{(1)} = \gamma_1\Lambda^{(2)}, 
\label{2storder_nz} \\
\left ( n_x^{(1)} \right )^2 + \left ( n_y^{(1)} \right )^2 + 2 n_z^{(2)} = 0,
\label{2ndorder_NORM} \\
\partial_z^2 u_x^{(2)} = 0,
\label{2storder_ux} \\
\partial_z^2 u_y^{(2)} = 0.
  \label{2storder_uy}  
\end{eqnarray}
Considering vanishing boundary conditions at the confining surfaces and \eqref{1storder_nz}, the only non-trivial solutions of the above system are 
\begin{equation}
  n_z^{(2)}(z) = - \frac{\left ( n_x^{(1)} \right )^2 + \left ( n_y^{(1)} \right )^2 }{2},  
\end{equation}
and
\begin{equation}
 \Lambda^{(2)}(z) = \gamma_1^{-1} \Bigl ( \partial_z^2 n_z^{(2)} - \alpha_3 L_z^{-1} n_x^{(1)} \Bigr ).
\end{equation}
$n_z^{(2)}(z)$ is shown in Fig.~\ref{SI-fig_nz2} for specific values of $\alpha_2$ and $\alpha_3$ for several ratios of the pitch over the cell thickness $P/L_z$.

\begin{figure}[th]
\center
\includegraphics[width=0.49\linewidth]{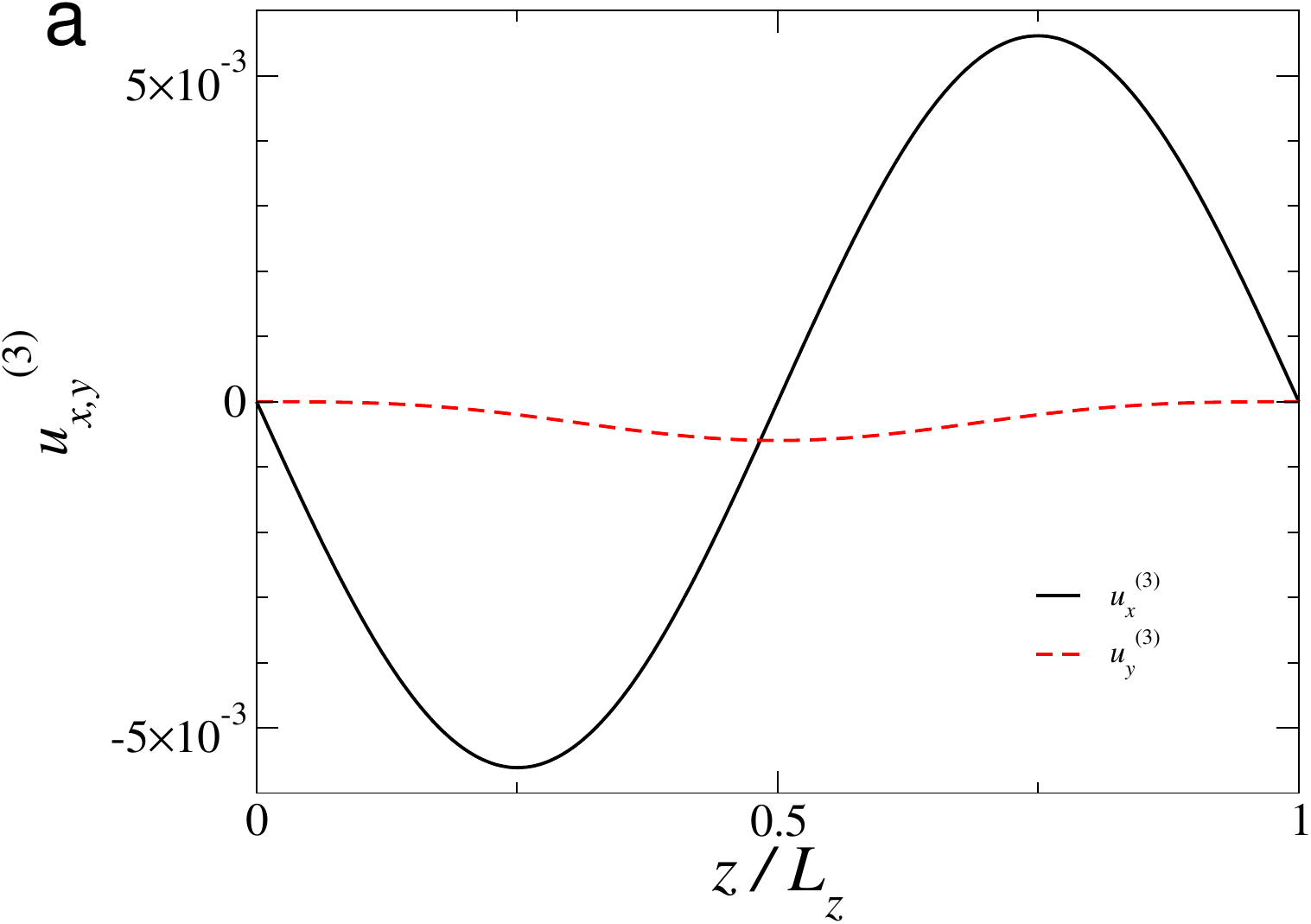}
\includegraphics[width=0.49\linewidth]{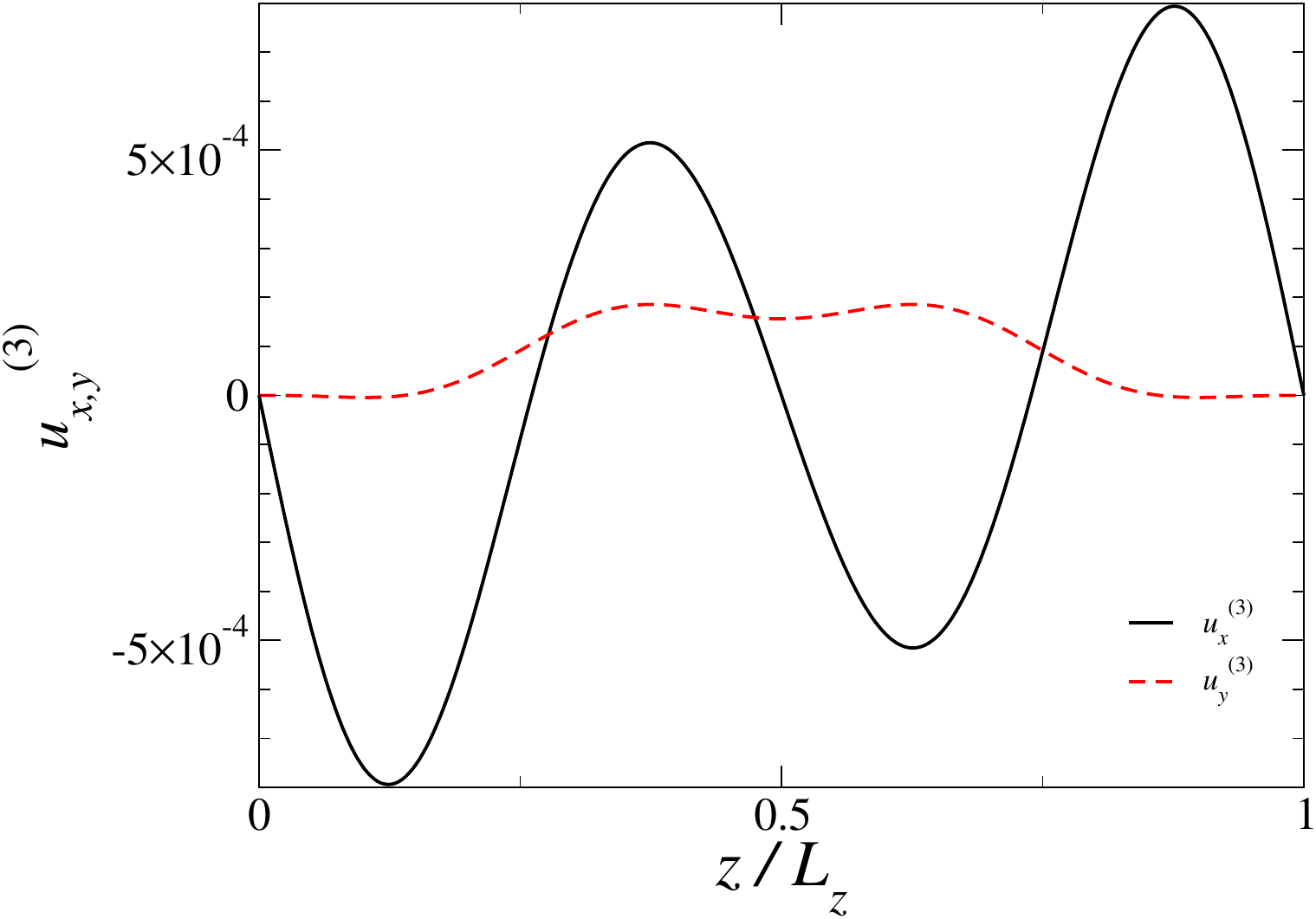}
\caption{${\cal O} (\varepsilon^3)$ contributions $u_x^{(3)}(z)$ and $u_y^{(3)}(z)$ to the flow field, obtained from \eqref{3storder_ux} and \eqref{3storder_uy}. The curves are computed at (a) $P/L_z = 2.5$, and (b) $ P/L_z = 1.25$, where $P$ is the cholesteric pitch and $L_z$ is the thickness of the LC cell. %and with the Leslie viscosities: $\alpha_4=0.08$ Pa.s, $\alpha_1 = 0.08 \alpha_4$, $\alpha_2 = -0.93\alpha_4$, $\alpha_3 = -0.014 \alpha_4$, $\alpha_5 = 0.56 \alpha_4 $ and $\alpha_6 = -0.41 \alpha_4$, which corresponds to a typical LC such as MBBA at 22$^\circ$C~\cite{de1993physics}.
} 
\label{SI-fig_Uxy3}
\end{figure}

\begin{figure}[th]
\center
\includegraphics[width=0.49\linewidth]{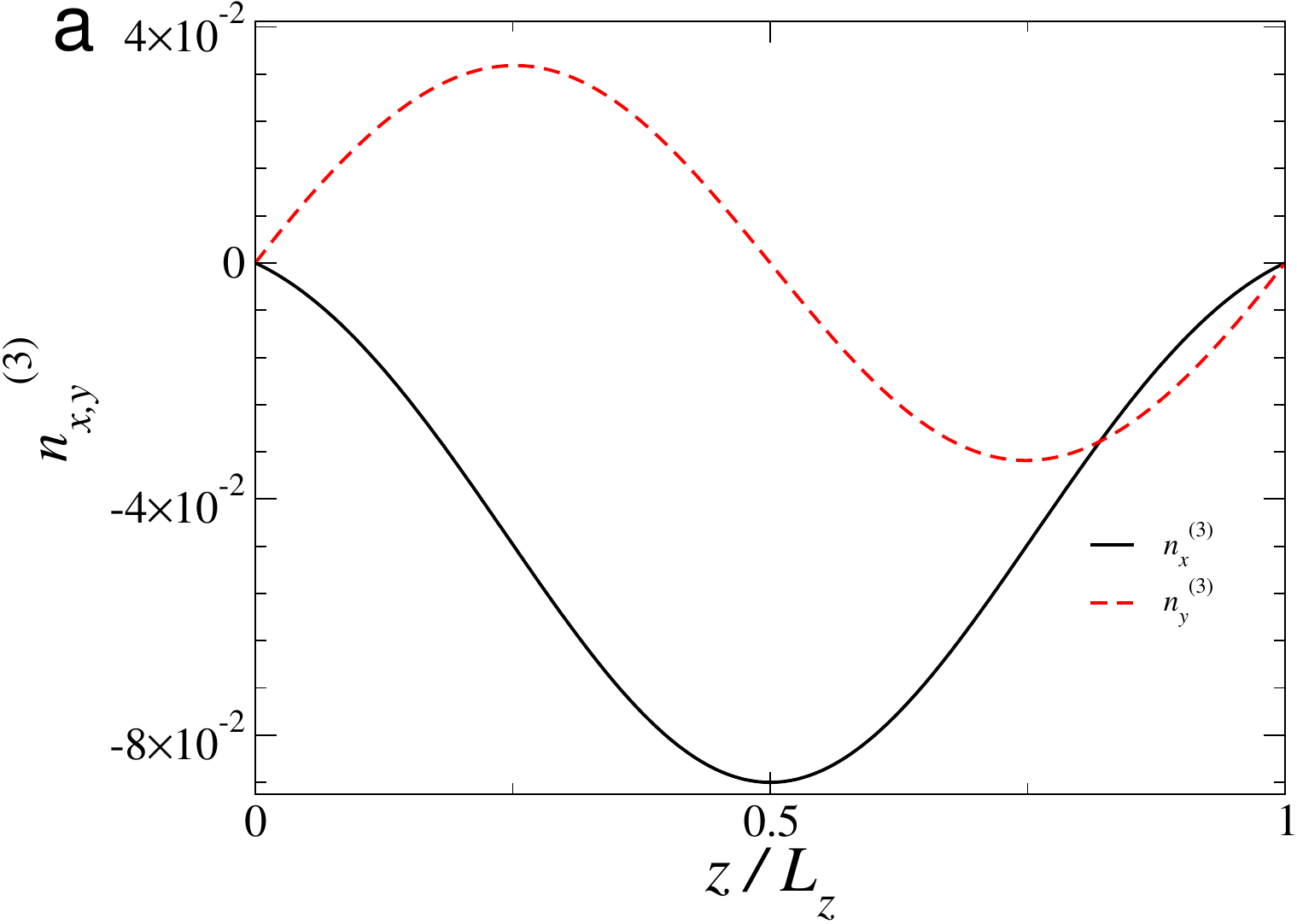}
\includegraphics[width=0.49\linewidth]{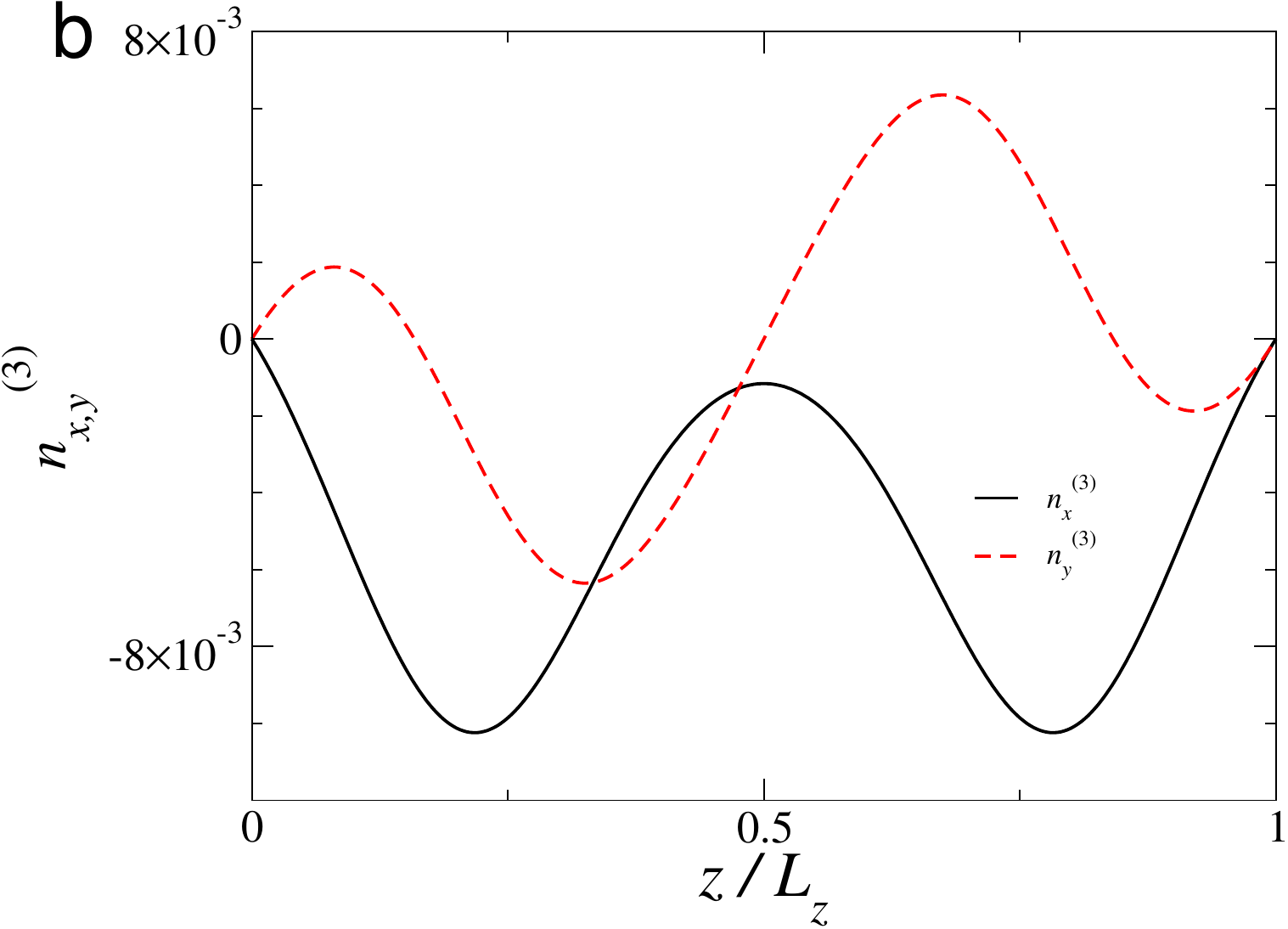}
\caption{${\cal O} (\varepsilon^3)$ contributions $n_x^{(3)}(z)$ and $n_y^{(3)}(z)$ to the director distortions, obtained from  \eqref{3storder_ux}, \eqref{3storder_uy} and 
\eqref{3storder_nx}, \eqref{3storder_ny}. The curves are computed at (a) $P/L_z = 2.5$, and (b) $ P/L_z = 1.25$, where $P$ is the cholesteric pitch and $L_z$ is the thickness of the LC cell.
% and with the Leslie viscosities: $\alpha_4=0.08$ Pa.s, $\alpha_1 = 0.08 \alpha_4$, $\alpha_2 = -0.93\alpha_4$, $\alpha_3 = -0.014 \alpha_4$, $\alpha_5 = 0.56 \alpha_4 $ and $\alpha_6 = -0.41 \alpha_4$, which corresponds to a typical LC such as MBBA at 22$^\circ$C~\cite{de1993physics}.
} 
\label{SI-fig_nxy3}
\end{figure}

Finally, collecting the terms ${\cal O}(\varepsilon^3)$ renders

\begin{eqnarray}
\partial_z^2 n_x^{(3)} + 2\partial_z n_y^{(3)} - \alpha_2 \left ( \partial_z u_x^{(3)} + n_z^{(2)}L_z^{-1} \right ) = \gamma_1 \left ( \Lambda^{(1)}n_x^{(2)} + \Lambda^{(2)}n_x^{(1)} \right ),
\label{3storder_nx} \\
\partial_z^2 n_y^{(3)} - 2\partial_z n_x^{(3)} - \alpha_2\partial_z u_y^{(3)} = \gamma_1 \left ( \Lambda^{(1)}n_y^{(2)} + \Lambda^{(2)}n_y^{(1)} \right ), 
\label{3storder_ny} \\
\partial_z^2 n_z^{(3)}  = \gamma_1 \left ( \Lambda^{(1)}n_z^{(2)} + \Lambda^{(3)}\right ), 
\label{3storder_nz} \\
n_x^{(1)}n_x^{(2)} + n_y^{(1)}n_y^{(2)} + n_z^{(3)} = 0,
\label{3storder_NORM} \\
\partial_z^2 u_x^{(3)} + {\cal A}L_z^{-1}\partial_z \left (n_x^{(1)}n_x^{(1)} \right ) + {\cal B} L_z^{-1}\partial_z n_z^{(2)}= 0,
\label{3storder_ux} \\
\partial_z^2 u_y^{(3)} + {\cal A}L_z^{-1} \partial_z \left (n_x^{(1)}n_y^{(1)} \right ) = 0,
  \label{3storder_uy}  
  \end{eqnarray}
  where
  \begin{eqnarray}
      {\cal A} = \frac{2 \alpha_1 + \alpha_3 + \alpha_6}{1 + \alpha_5 - \alpha_2}, \\
      {\cal B} = \frac{2(\alpha_5 - \alpha_2)}{1 + \alpha_5 - \alpha_2}.
  \end{eqnarray}
Taking into account the solutions $n_x^{(2)}(z)=n_y^{(2)}(z)=0$, we find from \eqref{3storder_nz} that $n_z^{(3)}(z) = 0$ (considering the vanishing boundary conditions at both surfaces), and next from \eqref{3storder_NORM} it follows $\Lambda^{(3)}(z)=0$. The 3rd order corrections to the flow field can be written as 
\begin{eqnarray}
%    u_x^{(3)}(z) = \frac{1}{384 L^3}\Biggl[ 24 {\cal B}(2z -L)\frac{\alpha_2}{\alpha_3} + L \biggl ( %2 \cot(L) (L-2z)\Bigl ( 9{\cal A} + 8 {\cal B}z(z-L)\frac{\alpha_2}{\alpha_3}\Bigr ) +\nonumber \\
%    6{\cal A} L \cos[4z]\cot[L] + 24\Bigl(\cos[2z] - \cot[L]\sin[2z]\Bigr)\Bigl( {\cal B}\frac{\alpha_2}{\alpha_3} -{\cal A}L\cot[L] \Bigr ) - \nonumber \\
%     3{\cal A}L\cos[2L] \csc^2[L]\sin[4z]
%    \biggr ) \Biggr ],
    u_x^{(3)}(z) = \frac{\alpha_2^2}{384 L_z^3}\Biggl( 4{\cal B}(L_z-2z)\Bigl (2z(L_z-z) -3 \Bigr ) + 3 L_z \csc[L_z]^2 \nonumber \\
     \biggl ( -2{\cal B}\cos[2(L_z -z)] + 3{\cal A}(L_z-2z )\sin[2L_z] + {\cal A}L_z\Bigl [ \sin[2(L_z -2z)] - \nonumber \\ 
     8\cos[L_z]\sin[L_z-2z] \Bigr] +2{\cal B}\Bigl[ \cos[2z] + 4z\sin[L_z]\cos[L_z-2z] -2 L_z\sin[2z]\Bigr] \biggr ) \Biggr ),
\end{eqnarray} 
and
\begin{eqnarray}
    u_y^{(3)}(z) = \frac{{\cal A} \alpha_2^2 \csc[L]}{128 L^2} \Biggl (  4 \cos[L] \Bigl ( 2z(L_z - z) - 1\Bigr ) + 4\cos[L_z-2z] + \nonumber \\
     \csc[L_z]\biggl ( 3L_z\cos[2L_z] +L_z \cos[2(L_z-2z)] + 4(z-L_z)\cos[ 2(L_z-z)] -4 z\cos[2z]\biggr )  \Biggr ).
\end{eqnarray}
$u_x^{(3)}(z)$ and $u_y^{(3)}(z)$ are plotted in Fig.~\ref{SI-fig_Uxy3} for specific values of the Leslie viscosities, and Fig.~\ref{SI-fig_nxy3} depicts $n_x^{(3)}(z)$ and $n_y^{(3)}(z)$.

% Bibliography
%\bibliography{ref} 

%Control: production of eprint (1) enabled
%

\newpage

\section{supplemental figures}

%-----------------figure-------------------------------
%------------------------------------------------------

%-----------------figure-------------------------------
%------------------------------------------------------

%-----------------figure-------------------------------
%------------------------------------------------------

%-----------------figure-------------------------------
%------------------------------------------------------

\begin{figure}[htb]
\includegraphics[width=0.65\textwidth]{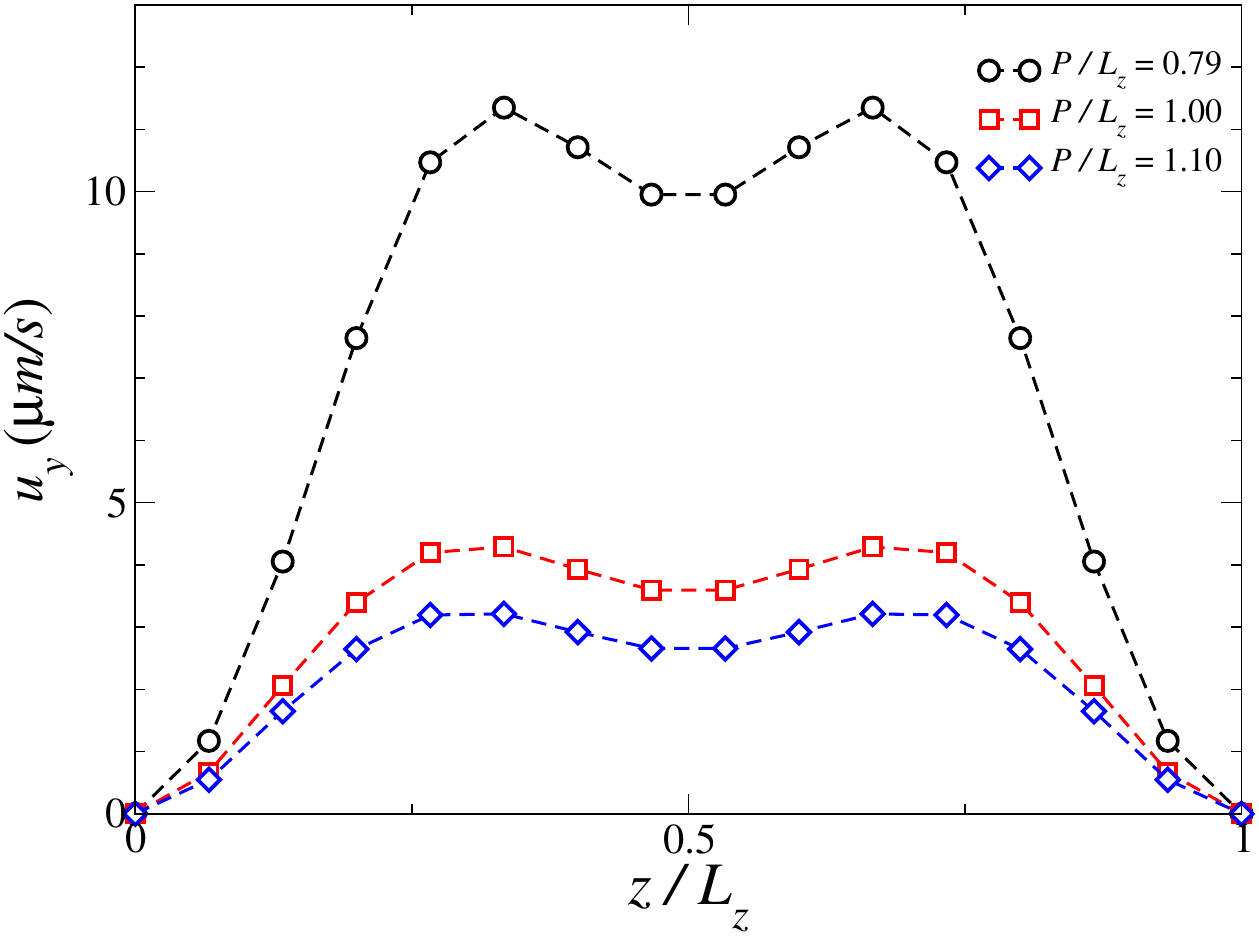}
\caption{The $y$- component $u_y$ of the steady state flow velocity as a function of the distance $z$ from the lower plate for the case of no toron. The results are from the Lattice Boltzmann simulations of the Ericksen-Leslie equations (1)-(11) in the main text. The Ericksen number $Er = 11.20$.}
%, where $\langle u \rangle$ is the space average magnitude of the fluid velocity at the steady state, $\alpha_4$ is the Leslie viscosity, $L_z$ the cell width and $K_{11}$ the splay elastic constant.}
%(b) Average over $z$ of $u_y(t,z)$ as a function of time, obtained from Lattice Boltzmann simulations of the Ericksen-Leslie equations. Results for systems with and without a toron are shown. Flow fields generated by a liquid crystal with opposite chiralities is shown (no toron present) for Ericksen number $Er = 1.54$. }
\label{fig:LB_on_u_y}
\end{figure}

%-----------------figure-------------------------------
\begin{figure}[th]
\center
\includegraphics[width=0.65\linewidth]{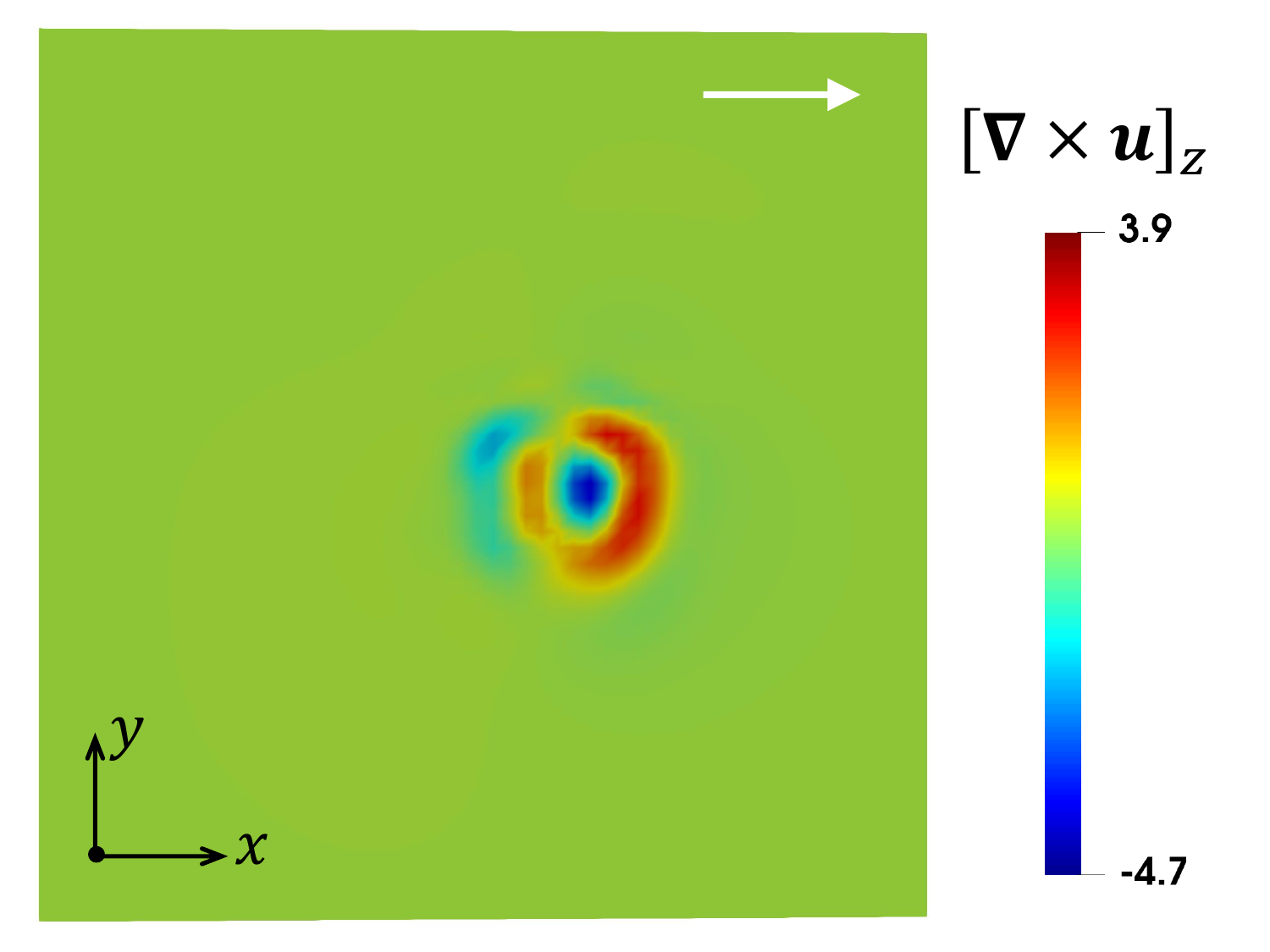}
\caption{Cross sectional view, in the plane $z=L_z/2$, of the $z$-component of the vorticity $[\nabla\cross \uvec]_z$, corresponding to the steady state solution of the Ericksen-Leslie equations. The model parameters are the same as in main figure 7. The white arrow indicate the shear direction.} 
\label{SI-fig_vorticity}
\end{figure}
%------------------------------------------------------

%-----------------figure-------------------------------
\begin{figure}[th]
\center
\includegraphics[width=0.65\linewidth]{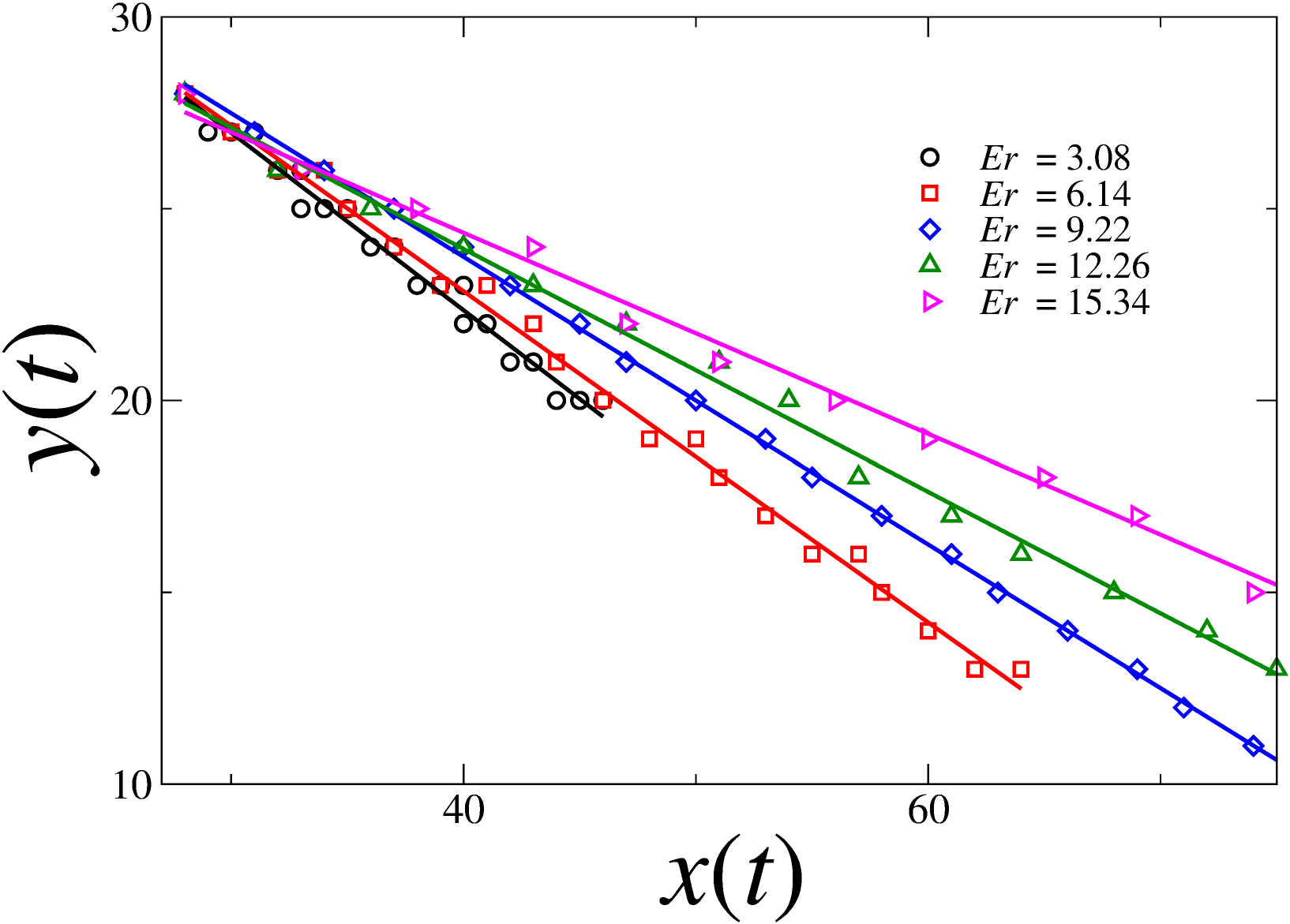}
\caption{Parametric plots $(x(t),y(t))$ of the toron trajectories, where $t$ is time, for several values of the Ericksen number, as obtained from the lattice Boltzmann simulations. Solid lines are linear fits to the data points (symbols). The curves correspond to the values of the Hall angles $\Theta_{HA}$ shown in  main figure 8(i).} 
\label{SI-fig_trajectories}
\end{figure}
%------------------------------------------------------